\documentclass[twocolumn,tighten,times]{aastex631}
\usepackage{mathrsfs}
\usepackage{longtable}
\usepackage{amsmath}
\usepackage{url}
\usepackage{mathtools}
\usepackage[normalem]{ulem}
\usepackage{graphicx}
\usepackage{float}
\usepackage{xcolor}
\usepackage{apjfonts} 
\turnoffedit
\usepackage{mathrsfs}
\usepackage{xspace}
\bibpunct{(}{)}{;}{a}{}{,}
\usepackage{amsmath}
\usepackage{natbib}
\usepackage{color}
\bibpunct{(}{)}{;}{a}{}{,}

\newcommand\aproxgt{\mathrel{%
      \rlap{\raise 0.511ex \hbox{$>$}}{\lower 0.511ex \hbox{$\sim$}}}}
\newcommand\aproxlt{\mathrel{%
      \rlap{\raise 0.511ex \hbox{$<$}}{\lower 0.511ex \hbox{$\sim$}}}}
\newcommand{\ignore}[1]{}
\newcommand{\storm}{{STORM}}
\newcommand{\civ}{\mbox{\rm C\,{\sc iv}}}

\newcommand{\hi}{\mbox{\rm H\,{\sc i}}}

\newcommand{\heii}{\mbox{\rm He\,{\sc ii}}}
\newcommand{\nv}{\mbox{\rm N\,{\sc v}}}
\newcommand{\ovi}{\mbox{\rm O\,{\sc vi}}}
\newcommand{\ovii}{\mbox{\rm O\,{\sc vii}}}
\newcommand{\siiv}{\mbox{\rm Si\,{\sc iv}}}
\newcommand{\svi}{\mbox{\rm S\,{\sc vi}}}
\newcommand{\lya}{\mbox{\rm Ly{$\alpha$}}}

\newcommand{\lyg}{\mbox{\rm Ly{$\gamma$}}}
\newcommand{\ciii}{\mbox{\rm C\,{\sc iii}}}
\newcommand{\cvi}{\mbox{\rm C\,{\sc vi}}}

\DeclareUnicodeCharacter{2212}{-}

\begin{document}

\title{AGN STORM 2: VIII. Investigating the Narrow Absorption Lines in Mrk~817 Using HST-COS Observations\footnote{Based on observations made with the NASA/ESA Hubble Space Telescope, obtained at the Space Telescope Science Institute, which is operated by the Association of Universities for Research in Astronomy, Inc., under NASA contract NAS5-26555. These observations are associated with program GO-16196, and archival data from programs 11505, 11524, and 17105.}}

\author[0000-0002-0964-7500]{Maryam~Dehghanian}
\affiliation{Department of Physics, Virginia Tech, Blacksburg, VA 24061, USA}

\author[0000-0003-2991-4618]{Nahum~Arav}
\affiliation{Department of Physics, Virginia Tech, Blacksburg, VA 24061, USA}

\author[0000-0002-2180-8266]{Gerard A.\ Kriss}
\affiliation{Space Telescope Science Institute, 3700 San Martin Drive, Baltimore, MD 21218, USA}

\author[0000-0002-4994-4664]{Missagh Mehdipour}
\affiliation{Space Telescope Science Institute, 3700 San Martin Drive, Baltimore, MD 21218, USA}

\author[0000-0002-3687-6552]{Doyee~Byun}
\affiliation{Department of Physics, Virginia Tech, Blacksburg, VA 24061, USA}

\author[0000-0001-6421-2449]{Gwen~Walker}
\affiliation{Department of Physics, Virginia Tech, Blacksburg, VA 24061, USA}

\author[0009-0001-5990-5790]{Mayank~Sharma}
\affiliation{Department of Physics, Virginia Tech, Blacksburg, VA 24061, USA}


\author[0000-0002-3026-0562]{Aaron J.\ Barth}
\affiliation{Department of Physics and Astronomy, 4129 Frederick Reines Hall, University of California, Irvine, CA, 92697-4575, USA}

\author[0000-0002-2816-5398]{Misty C.\ Bentz}
\affiliation{Department of Physics and Astronomy, Georgia State University, 25 Park Place, Suite 605, Atlanta, GA 30303, USA}

\author[0000-0001-6301-570X]{Benjamin D. Boizelle}
\affiliation{Department of Physics and Astronomy, N284 ESC, Brigham Young University, Provo, UT, 84602, USA}


\author[0000-0002-1207-0909]{Michael S.\ Brotherton}
\affiliation{Department of Physics and Astronomy, University of Wyoming, Laramie, WY 82071, USA}


\author[0000-0002-8294-9281]{Edward M.\ Cackett}
\affiliation{Department of Physics and Astronomy, Wayne State University, 666 W.\ Hancock St, Detroit, MI, 48201, USA}

\author[0000-0001-9931-8681]{Elena Dalla Bont\`{a}}
\affiliation{Dipartimento di Fisica e Astronomia ``G.\  Galilei,'' Universit\'{a} di Padova, Vicolo dell'Osservatorio 3, I-35122 Padova, Italy}
\affiliation{INAF-Osservatorio Astronomico di Padova, Vicolo dell'Osservatorio 5 I-35122, Padova, Italy}
\affiliation{Jeremiah Horrocks Institute, University of Central Lancashire, Preston, PR1 2HE, UK}

\author[0000-0003-3242-7052]{Gisella~De~Rosa}
\affiliation{Space Telescope Science Institute, 3700 San Martin Drive, Baltimore, MD 21218, USA}


\author[0000-0003-4503-6333]{Gary J.\ Ferland}
\affiliation{Department of Physics and Astronomy, The University of Kentucky, Lexington, KY 40506, USA}


\author{Carina Fian}
\affiliation{Departamento de Astronomía y Astrofísica, Universidad de Valencia, E-46100 Burjassot, Valencia, Spain}

\author{Alexei V.\ Filippenko}
\affiliation{Department of Astronomy, University of California, Berkeley, CA 94720-3411, USA}


\author[0000-0001-9092-8619]{Jonathan Gelbord}
\affiliation{Spectral Sciences Inc.
30 Fourth Ave, Suite 2
Burlington, MA 01803, USA}





\author[0000-0002-2908-7360]{Michael R.\ Goad}
\affiliation{School of Physics and Astronomy, University of Leicester, University Road, Leicester, LE1 7RH, UK}

\author[0000-0003-1728-0304]{Keith Horne}
\affiliation{SUPA School of Physics and Astronomy, North Haugh, St.~Andrews, KY16~9SS, Scotland, UK}


\author[0000-0002-0957-7151]{Yasaman Homayouni}
\affiliation{Department of Astronomy and Astrophysics, The Pennsylvania State University, 525 Davey Laboratory, University Park, PA 16802, USA} \affiliation{Institute for Gravitation and the Cosmos, The Pennsylvania State University, University Park, PA 16802, USA}
\affiliation{Space Telescope Science Institute, 3700 San Martin Drive, Baltimore, MD 21218, USA}

\author[0000-0002-1134-4015]{Dragana Ili\'{c}}
\affiliation{University of Belgrade - Faculty of Mathematics, Department of Astronomy, Studentski trg 16, 11000 Belgrade, Serbia}
\affiliation{Hamburger Sternwarte, Universit\"at Hamburg, Gojenbergsweg 112, 21029 Hamburg, Germany}

\author[0000-0003-0634-8449]{Michael D.\ Joner}
\affiliation{Department of Physics and Astronomy, N283 ESC, Brigham Young University, Provo, UT 84602, USA}


\author[0000-0003-0172-0854]{Erin A.\ Kara}
\affiliation{MIT Kavli Institute for Astrophysics and Space Research, Massachusetts Institute of Technology, Cambridge, MA 02139, USA}

\author[0000-0002-9925-534X]{Shai Kaspi}
\affiliation{School of Physics and Astronomy and Wise Observatory, Tel Aviv University, Tel Aviv 69978, Israel}

\author[0000-0001-6017-2961]{Christopher S.\ Kochanek}
\affiliation{Department of Astronomy, The Ohio State University, 140 W.\ 18th Ave., Columbus, OH 43210, USA}
\affiliation{Center for Cosmology and AstroParticle Physics, The Ohio State University, 191 West Woodruff Ave., Columbus, OH 43210, USA}
\author[0000-0003-0944-1008]{Kirk T. Korista}
\affiliation{Department of Physics, Western Michigan University, 1120 Everett Tower, Kalamazoo, MI 49008-5252, USA}
\author{Peter Kosec}
\affiliation{MIT Kavli Institute for Astrophysics and Space Research, Massachusetts Institute of Technology, Cambridge, MA 02139, USA}
\affiliation{Center for Astrophysics—Harvard $\&$ Smithsonian, Cambridge, MA, USA}

\author[0000-0001-5139-1978]{Andjelka B. Kova\v cevi\'c}
\affiliation{University of Belgrade - Faculty of Mathematics, Department of Astronomy, Studentski trg 16, 11000 Belgrade, Serbia}

\author{Hermine Landt}
\affiliation{Centre for Extragalactic Astronomy, Department of Physics, Durham University, South Road, Durham DH1 3LE, UK}

\author[0000-0002-8671-1190]{Collin Lewin}
\affiliation{MIT Kavli Institute for Astrophysics and Space Research, MIT, 77 Massachusetts Avenue, Cambridge, MA 02139, USA}








\author{Ethan R. Partington}
\affiliation{Department of Physics and Astronomy, Wayne State University, 666W. Hancock Street, Detroit, MI 48201, USA}



\author[0000-0003-2398-7664]{Luka \v{C}.\ Popovi\'{c}}
\affiliation{Astronomical Observatory Belgrade, Volgina 7, 11000 Belgrade, Serbia}
\affiliation{University of Belgrade - Faculty of Mathematics, Department of Astronomy, Studentski trg 16, 11000 Belgrade, Serbia}

\author[0000-0002-6336-5125]{Daniel Proga}
\affiliation{Department of Physics \& Astronomy, 
University of Nevada, Las Vegas 
4505 S.\ Maryland Pkwy, 
Las Vegas, NV, 89154-4002, USA}


\author{Daniele Rogantini}  
\affiliation{MIT Kavli Institute for Astrophysics and Space Research, Massachusetts Institute of Technology, Cambridge, MA 02139, USA}



\author{Matthew R. Siebert}  
\affiliation{Space Telescope Science Institute, 3700 San Martin Drive, Baltimore, MD 21218-2410, USA}

\author[0000-0003-1772-0023]{Thaisa Storchi-Bergmann}
\affiliation{Departamento de Astronomia - IF, Universidade Federal do Rio Grande do Sul, CP 15051, 91501-970 Porto Alegre, RS, Brazil}




\author[0000-0001-9191-9837]{Marianne Vestergaard}
\affiliation{Steward Observatory, University of Arizona, 933 North Cherry Avenue, Tucson, AZ 85721, USA}
\affiliation{DARK, The Niels Bohr Institute, University of Copenhagen, Jagtvej 155, DK-2200 Copenhagen, Denmark}



\author[0000-0002-5205-9472]{Timothy. Waters}
\affiliation{X-Computational Physics Division, Los Alamos National Laboratory,NM 87545, USA}


\author[0000-0001-9449-9268]{Jian-Min Wang}
\affiliation{Key Laboratory for Particle Astrophysics, Institute of High Energy Physics, Chinese Academy of Sciences, 19B Yuquan Road,\\ Beijing 100049, People's Republic of China}
\affiliation{School of Astronomy and Space Sciences, University of Chinese Academy of Sciences, 19A Yuquan Road, Beijing 100049, People's Republic of China}
\affiliation{National Astronomical Observatories of China, 20A Datun Road, Beijing 100020, People's Republic of China}



\author{Fatima Zaidouni}
\affiliation{MIT Kavli Institute for Astrophysics and Space Research, Massachusetts Institute of Technology, Cambridge, MA 02139, USA}

\author[0000-0001-6966-6925]{Ying Zu}
\affiliation{Tsung-Dao Lee Institute, Shanghai Jiao Tong University, Shanghai 200240, China}
\affiliation{Department of Astronomy, School of Physics and Astronomy, Shanghai Jiao Tong University, 800 Dongchuan Road, Shanghai, 200240, People's Republic of China}
\affiliation{Shanghai Key Laboratory for Particle Physics and Cosmology, Shanghai Jiao Tong University, Shanghai 200240, People's Republic of China}

\begin{abstract}
We observed the Seyfert 1 galaxy Mrk~817 during an intensive multi-wavelength reverberation mapping campaign for 16 months. Here, we examine the behavior of narrow UV absorption lines seen in HST/COS spectra, both during the campaign and in other epochs extending over 14 years.  We conclude that while the narrow absorption outflow system (at $-3750$ km~s$^{-1}$ with FWHM=177 km~s$^{-1}$) responds to the variations of the UV continuum as modified by the X-ray obscurer, its total column density (log $N_{\rm H}$ =19.5 $^{+0.61}_{-0.13}$ cm$^{−2}$)  did not change across all epochs. The adjusted ionization parameter (scaled with respect to the variations in the Hydrogen ionizing continuum flux) is log $U_{\rm H} =-1.0$ $^{+0.1}_{-0.3}$. The outflow is located at a distance smaller than 38 parsecs from the central source, which implies a hydrogen density of $n_{\rm H} > 3000$~cm$^{-3}$. The absorption outflow system only covers the continuum emission source and not the broad emission line region, which suggests that its transverse size is small (<~10$^{16}$ cm), with potential cloud geometries ranging from spherical to elongated along the line of sight.
\end{abstract}


\section{INTRODUCTION } \label{sec:intro}
Active galactic nuclei (AGNs) stand out as important tools for understanding the evolution of galaxies. In particular, AGN feedback
uses the deposition of energy and momentum into the
host's interstellar medium (ISM) to help regulate the star formation \citep[e.g.,][]{elvis06}. One possible contributor to the feedback process
is outflows detected as absorption features in
AGN spectra. \citep[e.g.,][]{silk98, scan04, yuan18, vayner21, he22}. These outflows provide a valuable understanding of the dynamics and physical processes occurring within the vicinity of SMBHs at the centers
of galaxies.

Absorption lines observed in the rest-frame UV spectra of AGNs are commonly classified into three categories: broad absorption lines (BALs), characterized by a width of $\geq$2000 kms$^{-1}$; narrow absorption lines (NALs), with a width of $\leq$500 kms$^{-1}$; and an intermediate group referred to as mini-BALs \citep{itoh20}.

Because of the difficulty in distinguishing between intrinsic NALs associated with the quasars and NALs that are unrelated to the quasars (intervening NALs), our understanding of the nature of NAL outflows is limited.
Intervening NALs can have diverse origins, including intervening galaxies, intergalactic clouds, Milky Way gas, or gas within the host galaxies of the quasars \citep{misa07a}. Various studies, such as \cite{misa07a, chen13}, and \cite{chen18a}, emphasize that observing variability in the absorption lines is a reliable method to differentiate between NALs originating from an associated outflow and those classified as intervening NALs. These variations in the absorption lines typically result from changes in the ionizing flux striking the absorbing gas.

While NAL outflows have garnered less attention compared to BALs, they may be a valuable tool for probing the physical properties of outflows (for example \cite{deh19} used NALs to explain the physics behind the line-continuum decorrelation observed in the Seyfert galaxy NGC 5548). This potential arises from two key reasons, as outlined by \cite{misa07}:

\begin{itemize}
\item NALs do not suffer from self-blending, a problem created by the merging of blue and red components of doublets like \civ\  $\lambda\lambda$1548,1551\AA. This simplifies the analysis of NALs, making them advantageous for certain investigations.

\item NALs are found in the spectra of both radio-loud and radio-quiet AGN,
whereas BALs are predominantly detected in radio-quiet
quasars. 
\end{itemize}

Approximately 50$\%$ of Seyfert galaxies and low-redshift AGNs exhibit intrinsic NALs \citep{cren99, cren03, dun08}. Relative to the emission lines, these narrow absorption lines are blue-shifted, commonly with outflow velocities around a few hundred km~s$^{-1}$.
However, higher-velocity components exceeding 1000 km~s$^{-1}$  are also observed in a few objects. Studies employing either variability \citep[e.g.,][]{gabel05,arav12,kriss19,arav20} or density-sensitive lines \citep[e.g.,][]{gabel05,arav15} locate the narrow absorption outflow gas in proximity to either the obscuring torus or the narrow-line region.
These outflow systems could originate from the obscuring torus \citep{krol95,krol01} or potentially from interstellar clouds in close proximity to the nucleus \citep{cren05}.

The lower-velocity lines exhibit physical characteristics typical of gas found in the narrow-line region or gas ablated from the torus, and their kinematics align with either a thermal wind originating from the torus or near-nuclear interstellar motions. For the case of the high-velocity lines that reach several thousand km~s$^{-1}$, an alternative acceleration mechanism is required. \cite{reva21} observed velocity ranges between 1000-2000 km~s$^{-1}$  in their study of nearby Seyferts' extended narrow-line regions, for which they suggest in situ radiative acceleration of existing clouds. Another possible scenario is that gas is shocked and entrained by higher-velocity outflows from the AGN itself, as proposed in entrained Ultra Fast Outflow (UFO) models \citep{gasp17,sanf18,sera19,long19,mehd22}.

\vspace{0.25cm}
\subsection{Seyfert Galaxy Mrk~817: The \storm2 Campaign}
The AGN \storm2\footnote{Space Telescope and Optical Reverberation Mapping~2 (the first STORM campaign targeted NGC 5548 \citep{derosa15}} project is an intensive spectroscopic reverberation mapping (RM) campaign that observed the Seyfert 1 galaxy Mrk~817 from 2020 to 2022 using the Cosmic Origins Spectrograph \citep[COS,][]{green12}, on the Hubble Space Telescope \citep[HST,][]{pet20}.  While the primary goal of the project was to determine the kinematics and geometry of the central regions using RM methods \citep{kara21,hom23a,part23,cac23,hom23b,neu23}, the observations revealed additional exciting results:   
there were significant variations in the response time of the broad UV emission lines to the continuum variations, and there was significant, variable absorption in the UV and soft X-rays. Specifically, broad emission lines such as \civ\  exhibited time lags ranging from 2 to 13 days during different time intervals. These variations are associated with variations in the characteristics of a UV and X-ray obscurer located between the broad line region and the central source \citep{hom23a}.

The significant difference observed in the response time of emission lines is a result of the variations in the properties of an X-ray obscurer \citep[e.g.,][]{kaast14}, which appears to be located between the Broad Line Region (BLR) and the central source \citep{hom23b}. The spectral energy distribution (SED) generated by the AGN must traverse the obscurer before reaching the BLR. \cite{kara21} (hereafter Paper I), show what the "obscured" SED looks like for a single visit; however, owing to the changing obscurer properties (location, column density, covering fraction, etc.), the obscured SED varies for each visit. \cite{part23} studied these obscurers using X-ray spectra from the Neutron Star Interior Composition ExploreR (NICER) aboard the International Space Station.

Paper~I modeled the high-velocity ($v_{\rm outflow}\approx$3720 km$^{-1}$) NAL in Mrk~817 for a single visit on December 2020 (called visit 3n), using both obscured and unobscured SEDs. They identified a set of narrow absorption lines (including \hi, \civ, \nv, \ovi, \siiv, and \svi) and used them to determine the photoionization structure of the absorption outflow system. Their results indicate that the absorbing gas is ionized by the obscured SED and has a hydrogen column density of $N_{\rm H}$=10$^{19.5}$ cm$^{-2}$,and an ionization parameter of log$\xi$=~1 erg~cm~s$^{-1}$, corresponding to log $U_{\rm H}$=$-0.25$ as explained below\footnote{The dimensionful ionization parameter $\xi$ is defined as 
$\xi= \frac{L}{n_{\rm H} R^2}$ \citep{Tarter69, Kallman01},
where $L$ is the ionizing luminosity and  $R$ is the distance from the source. The dimensionless ionization parameter $U_{\rm H}$ is defined in Equation~\ref{eq1}.}. 

In this study, we model the high-velocity narrow absorption lines observed in seven HST spectra, as listed in Table~\ref{tab1}. This includes the observation already modeled in Paper~I (visit 3n). Two of the observations date back to 2009; one was conducted in 2023, while the remaining four are from \storm2 observations conducted between 2020 and 2022. We selected these particular \storm2 observations from a pool of 165 available spectra because they are the only ones with such extended wavelength coverage, reaching down to 940~\AA, and so include the \ovi\  and \svi\  doublets and higher-order Lyman lines. This broad coverage enhances our ability to construct a well-constrained photoionization model.

We will examine how this high-velocity absorption component fits into the general population of narrow absorption lines in Seyferts. The \storm2 observations show that this absorption component became noticeable when the obscuring outflow appeared in Mrk~817. This mirrors the behavior of Component~1 in NGC 5548 (Arav et al. 2015; Dehghanian et al. 2019), which was characterized by a high velocity (1350 km~s$^{-1}$) and increased strength when an obscuring outflow appeared in NGC 5548. Both absorption systems are also seen in a variety of ionic species beyond \civ\ and \lya.

We investigate whether, in all cases, the NALs responded to the obscured SED. We then employ photoionization models to deduce the total hydrogen column density and the ionization parameters of the outflow system for each visit. This approach enables us to explore the potential variations of the absorption outflow system over almost 14 years. Our findings reveal that, while the absorption outflow system reacts to both AGN and obscurer variations, it remains notably stable over time. 

The structure of the paper is outlined as follows:
In Section~\ref{sec:obs}, we describe the observations and data acquisition of Mrk~817. In Section~\ref{sec:anal}, we describe the analysis and explain the methods used in the paper. Section~\ref{sec:anal} also details the methodologies used to calculate the ionic column densities of the NALs.  In Section~\ref{sec-photo}, we describe our photoionization models and report the results for each visit. Finally, Section~\ref{sec:dis} summarizes the paper and discusses the results.
\begin{table*}
\caption{HST/COS observations\label{tab1}}
 \centering
\setlength{\tabcolsep}{8pt} 
\renewcommand{\arraystretch}{1.5}
\begin{tabular}{c c c c c c} 
 \tableline
 Visit ID & THJD & Date & F1180 & Data Source & Grating \\ [0.8ex] 
 \hline
 \hline
09-1 &  5047.1 & 2009-08-04 & 8.22$\times 10^{-14}$ & COS-GTO (PID$^{a}$:~11505)&G130+G160\\
09-2 & 5193.4 & 2009-12-28 & 5.91$\times 10^{-14}$ & COS-GTO (PID:11524) & G130+G160\\
 \hline
 3n &  9202.3 & 2020-12-18 & 1.20$\times 10^{-13}$ & \storm2 (PID:16196) & G130(1096$^{b}$)+G160\\ 
 75 & 9322.4 & 2021-04-18 & 1.09$\times 10^{-13}$ & \storm2 (PID:16196) &  G130(1096)+G160\\
 2n & 9581.5 & 2022-01-02 & 7.95$\times 10^{-14}$ & \storm2 (PID:16196) & G130(1096)+G160 \\
 4d & 9634.3 & 2022-02-24 & 9.50$\times 10^{-14}$ & \storm2 (PID:16196) & G130(1096)+G160\\
 \hline
 A5  & 10129.8 & 2023-07-04 & 1.41$\times 10^{-13}$ & \cite{proptoo} (PID:17105) & G130(1096)+G130(1222)+G160 \\
\tableline
\end{tabular}
\tablecomments{Details of the "before \storm2" observations (first two rows), the "\storm2" observations (4 middle rows), and the "after \storm2" observations (last row). The observation times use “truncated
Heliocentric Julian Dates,” defined as THJD = HJD$-2450000$. F1180 is the continuum flux at $\lambda$1180\AA \ in erg cm$^{-2}$~s$^{-1}$~\AA$^{-1}$.
\\$a$: Proposal Identification Number
\\$b$: Central Wavlength in \AA}
\end{table*}
Here we adopt a cosmology with $h$= 0.696, $\Omega_{m}$= 0.286, and
$\Omega_\Lambda$ = 0.714 \citep{benn14}. 
\section{Observations}
\label{sec:obs}

Mrk~817 (PG 1434+590) is a Seyfert 1 galaxy with a systemic redshift of $z$=0.031455 \citep{stra88} and located at J2000 RA=14:36:22.08 and Dec=+58:47:39.39 (based on the NASA Extragalactic Database, NED).\footnote{NED:\url{https://ned.ipac.caltech.edu/}.NED is funded by the National Aeronautics and Space Administration and operated by the California Institute of Technology.}
\cite{falco99} later updated the above-mentioned redshift to be $z$ = 0.031158, meaning $\Delta cz$=89 km~s$^{-1}$, which translates to $<3\%$ decrease in our estimated outflow velocity.

The first series of AGN \storm2 observations targeted Mrk~817 for 165 epochs of HST visits. During these observations, which happened between Nov 2020 and Feb 2022, we used the COS instrument with G130M and G160M gratings to cover the
1070\AA\ – 1750\AA \ range in single-orbit visits with an approximately
2-day cadence. Extensive details of the observations can be found in \cite{hom23a}.
Table~\ref{tab1} summarizes the HST visits discussed in this paper. The visit labeled as "visit 3n" is the same spectrum that is modeled in Paper~I. Visits 09-1 and 09-2 are from proposals GO-11505 and GO-11524 performed by \cite{Pro09-1} and \cite{pro09-2}, respectively, and published by \cite{wint11}. These are "before \storm2" observations. 
Visits 3n, 75, 2n, and 4d are \storm2 observations \citep{pet20} and finally, visit A5 is an observation conducted by \cite{proptoo} "after \storm2" (program GO-17105).
We add that \citep{pent2000} discussed the local \lya\  forest in Mrk 817 using HST/GHRS data. Mrk~817 was also one of the early targets observed by the COS-GTO team, including the data listed in this paper as visits 09-1 and 09-2. 
\begin{figure*}
\centering
\includegraphics [width=\textwidth]{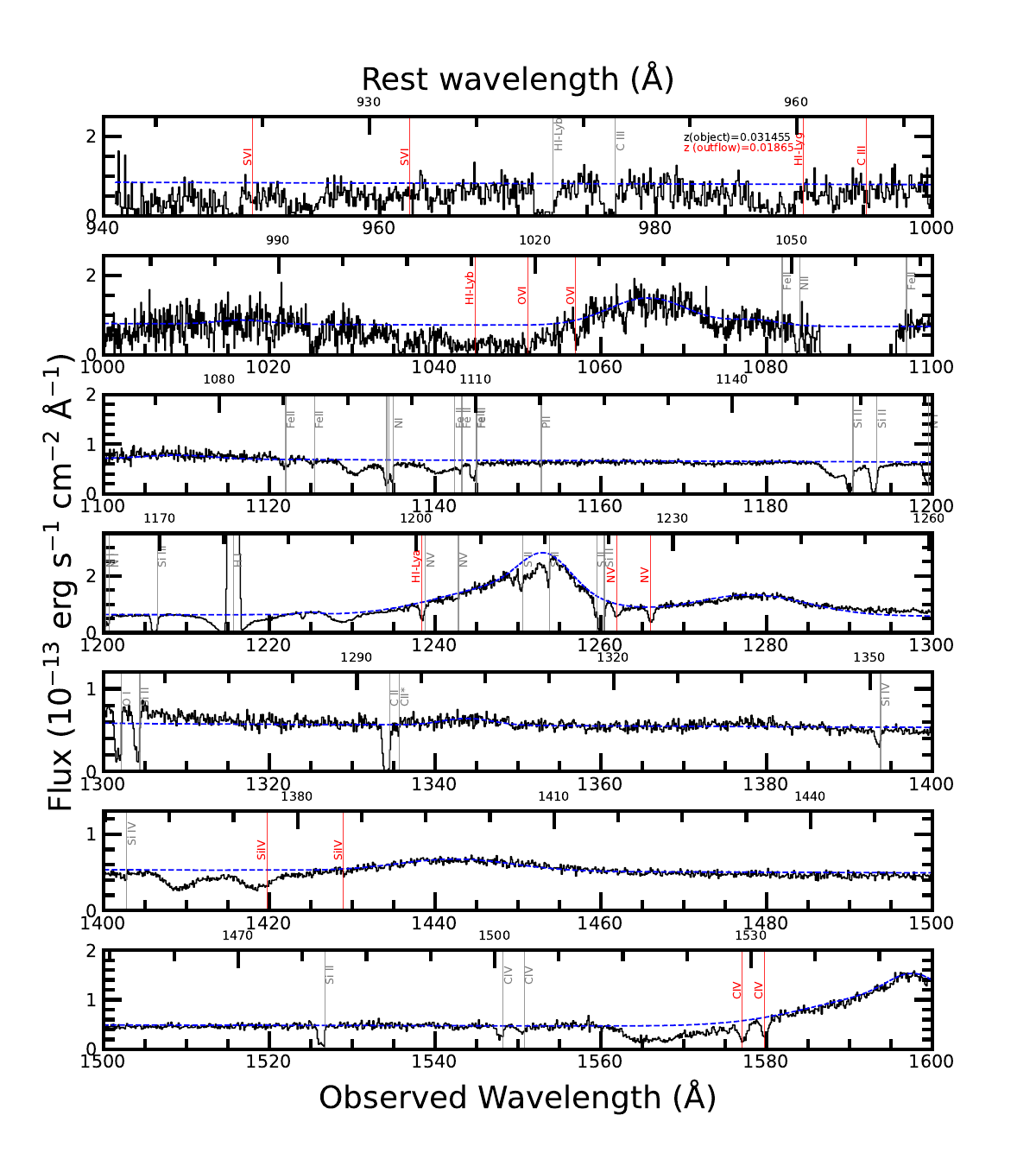}
 \caption{The January 2022 HST/COS spectrum of Mrk~817  (visit 2n). Red lines indicate the absorption features of the outflow system at a velocity of $-3750$ km~s$^{-1}$. Absorption from the Galactic interstellar medium (ISM) is shown with vertical grey lines. The Galactic absorption along this sight-line includes strong blueshifted components \citep{coll03, Fox23}. The dashed blue line shows our continuum plus broad line emission model. The spectrum also shows broader higher-velocity outflow systems such as the \civ\  trough between 1560-1570 \AA\  observed wavelength.  Details for that system are given in Paper~I.}\label{figflux}
\end{figure*}
\section{Analysis}
\label{sec:anal}
After obtaining the data from the Mikulski Archive for
Space Telescopes (MAST), we identified an absorption outflow system at a velocity of
$v_{\rm {centroid}}$= $-3750$ km s$^{−1}$ and with a FWHM of 177 km s$^{−1}$ (measured based on the \civ\  narrow absorption line in visit 3n), with blueshifted ionic absorption lines denoted by
red vertical lines in Figure~\ref{figflux}. Among the identified absorption
lines, several well-known resonance doublets, such as \civ\  and \nv,
\ are observed.  Figure~\ref{figflux} shows an example of such an identification. The spectrum shown in this figure belongs to visit 2n (January 2022). 
\subsection{Ionic Column Densities}
\label{ionCD}
As an essential step for comprehending the physical characteristics of the outflow system, we need to determine the ionic column densities ($N_{\rm ion}$) of the NALs. The most straightforward method for measuring column densities is called the apparent optical depth method (AOD), in which we assume a uniformly covered homogeneous source \citep{sava91}. In this method, the first assumption is that 
\begin{equation}
    I(\lambda)=I_0(\lambda)e^{-\tau(\lambda)} 
\end{equation}
\noindent where 
$I(\lambda)$ is the intensity, $I_0(\lambda)$ is the modeled intensity without absorption, and 
$\tau(\lambda)$ is the optical depth as a function of wavelength. The second assumption is the constant opacity of the absorbing material, so when expressing the optical depth as a function of outflow velocity, it is related to the column density per unit velocity $N(v)$ cm$^{-2}$ (km~s$^{-1})^{-1}$  \citep[see Equation~8 of][]{sava91} by\\
\begin{equation}
    \tau(v)=\frac{\pi e^2}{m_e c}f\lambda N(v)
\end{equation}
\noindent where $m_e$ is the electron mass, $e$ is the elementary charge, $f$ is the oscillator strength and $\lambda$ is the wavelength of the transition line, respectively. 
Because of possible saturation (see below), the  AOD method is employed to determine lower limits on $N_{\rm ion}$ for singlets, as well as upper limits for doublets when there are no observable absorption troughs.

In cases where multiple lines of the same ion and energy state are present, the partial covering (PC) method can be utilized. This assumes a homogeneous source that is partially covered by the outflow \citep{barl97, arav99a, arav99b}. When using the PC method, phenomena such as non-black saturation are taken into account since a velocity-dependent covering factor is deduced \citep{kool02}. For doublets with $f$ value ratio of 1:2, the covering fraction $C(v)$ and the optical depth $\tau(v)$ can both be calculated using  \citep{arav05} as
\begin{eqnarray}
    I_R(v)-[1-C(v)]=C(v)e^{-\tau(v)}
\end{eqnarray}
and
\begin{eqnarray}
    I_B(v)-[1-C(v)]=C(v)e^{-2\tau(v)}
\end{eqnarray}
\noindent where $I_R(v)$ is the normalized intensity of the red absorption feature, $I_B(v)$ is the normalized intensity of the blue absorption feature, and $\tau(v)$ is the optical depth of the red component. Whenever the PC method is used, the final result is a measurement rather than an upper or lower limit. For the doublets of \civ, \ovi, and \siiv\  we use an $f$ value ratio of 1:2, since this approximation is less than 2\% different from their actual $f$ values reported by \citep{mort03}.

A detailed comparison between the PC prediction and numerical calculations of the optical depth of a clumpy medium showed the PC method to be surprisingly accurate even when the clumps evolve in a turbulent flow \citep{tim17}. For a more detailed explanation of different methods used to calculate ionic column densities and for a deeper understanding of the underlying logic and mathematical aspects, please see \cite{barl97,arav99a,arav99b, kool02, arav05,borg12a,byun22b, byun22c} and \cite{deh23}. In the following three subsections, we separately explain how we dealt with the various spectra and what lines were identified in each.

\subsubsection{\storm2: visits 3n, 75, 2n, and 4d}
\label{stormanal}
These observations are part of the \storm2 project and were obtained between 2020 and 2022. For each individual spectrum, we identified the resonance doublets of the \civ, \siiv, \nv, \ovi, and \svi \ absorption lines, along with the \lya, \lyg \, and \ciii\  absorption lines. Figure~\ref{figflux} shows the spectrum and absorption lines for visit 2n. The other three spectra are very similar to the spectrum shown in Figure~\ref{figflux}.  
\par
\begin{figure}
\centering
\begin{minipage}{3.5 in}
\includegraphics [width=3.5 in]{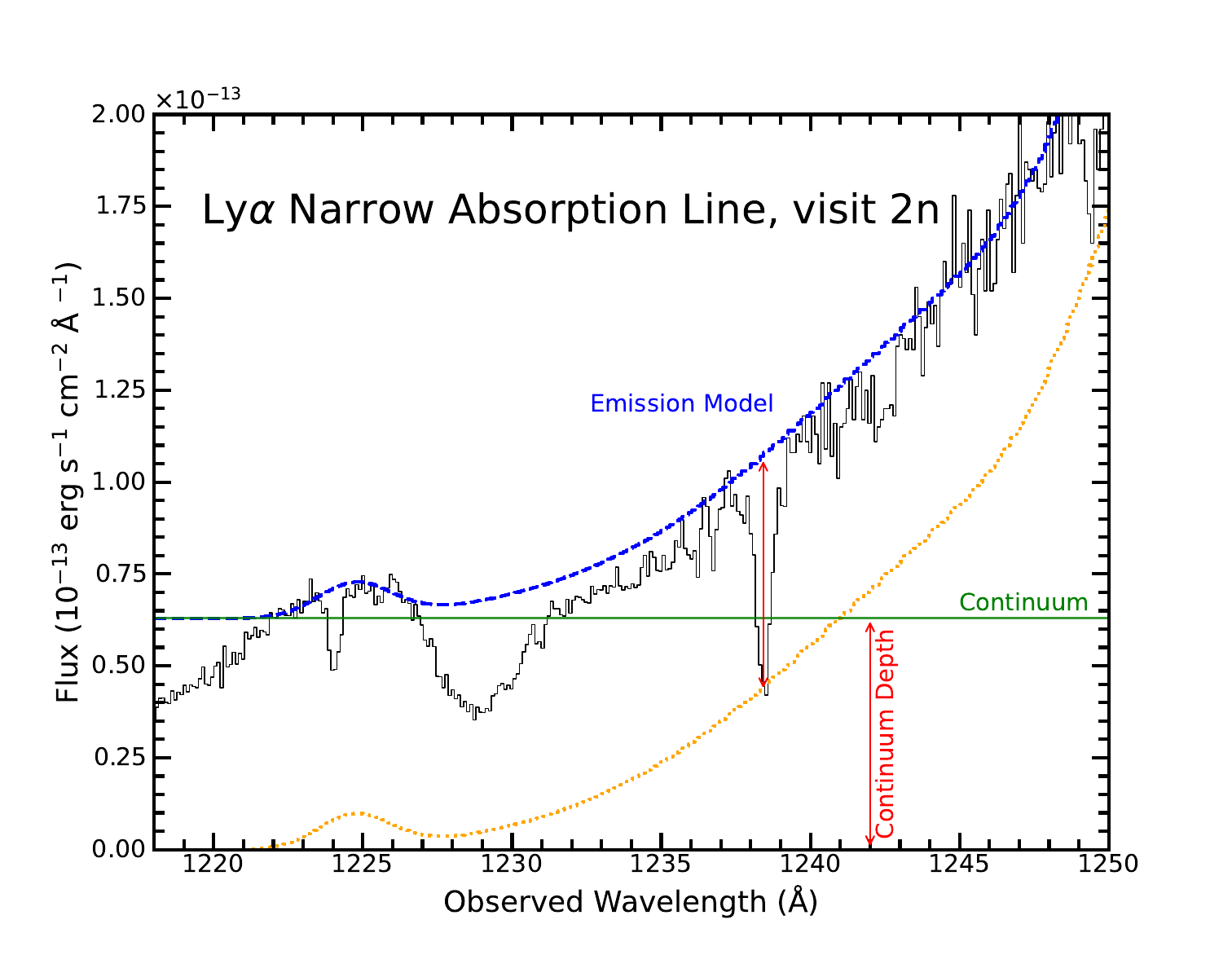}
\end{minipage}
 \caption{The visit 2n HST/COS spectrum of the Ly$\alpha$ surrounding region (black). The best-fit emission plus the continuum model is in blue; the model continuum is anchored in the line-free region at around 1200\AA\ and is shown in green. The orange line shows the continuum-subtracted emission model. Both of the red arrows have the same length, which is equal to the depth of the continuum. It is clear that the depth of the absorption trough is exactly equal to the continuum depth, indicating that it fully (and only, see section 3.1.1) covers the continuum emission.}\label{figlya}
\end{figure}

It is essential to highlight that, as explored in Paper~I by modeling the \nv\  and \ovi\ doublets, the narrow absorption outflow system primarily covers the continuum source emission rather than the BLR emission. Figure~\ref{figlya} confirms the same situation is happening for \lya. A detailed examination of the \civ\ NAL also supports the conclusion that its narrow absorption covers only the continuum and not the BLR. Based on these findings, it is appropriate to consider only the partial coverage of the continuum source and exclude any coverage of the BLR throughout this paper. This situation also occurs in some BAL outflows \citep[e.g., Figure 1c in][]{arav99b}. In Section~\ref{sec:dis}, we further discuss this by establishing an upper limit for the location of the outflow.

Figure~\ref{figlya} illustrates how we incorporate this assumption into our calculations. We use a power-law for the continuum and then fit the emission lines with Gaussian[s] to model the data. To produce Figure~\ref{figlya}, we subtract the continuum level from the emission model, resulting in an "emission-only" model (depicted by the orange curve in Figure~\ref{figlya}). Because the depth of the absorption line is equal to the subtracted continuum, we know that the absorber is only affecting the continuum and not the BLR emission. 
The same is true for the \nv \ and \ovi\  NALs (see Paper~I).  Combined, these independent observations strongly suggest that the outflow only covers the continuum source. Subsequently, we subtract this "emission-only" model (orange curve) from the total flux and model the narrow absorption lines under the assumption that they only cover the continuum emission and do not cover the BLR.

\begin{figure}
\centering
\begin{minipage}{3.5 in}
\includegraphics [width=3.5 in]{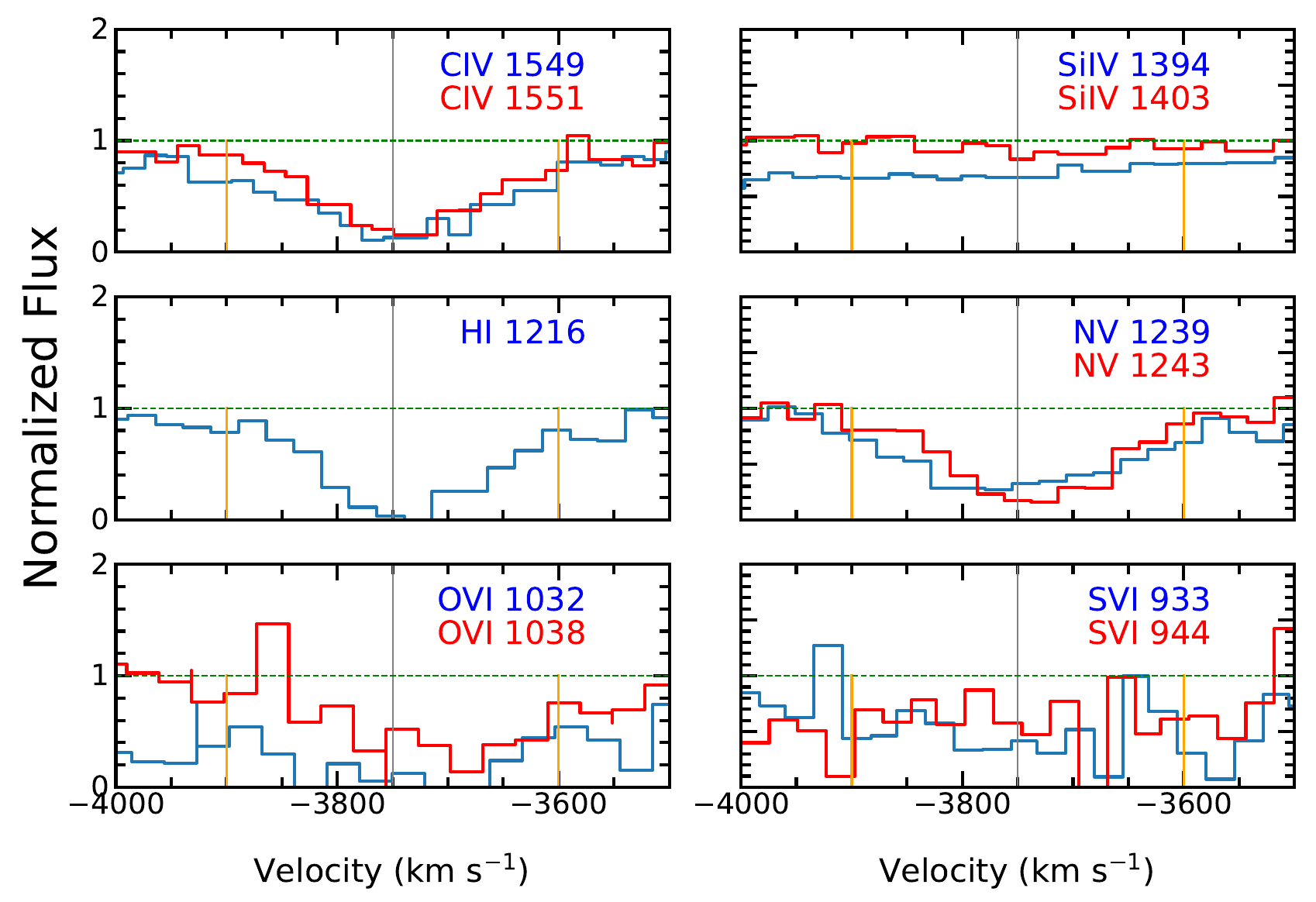}
\end{minipage}
 \caption{Normalized flux (after subtracting the emission lines, see Figure~\ref{figlya}) versus velocity for blueshifted absorption lines detected in the spectrum of Mrk~817 in Jan 2022 (visit 2n). The continuum level is shown by the horizontal green dashed line. The integration range ( $−3900$ to $−3600$
km s$^{−1}$)
is shown with vertical orange lines, while the centroid velocity of $v_{\textrm{centroid}}$ = $-3750$ kms$^{-1}$ is shown with a solid black line.}\label{figVel2}
\end{figure}

For each narrow absorption line, we transferred the emission-subtracted normalized flux from 
wavelength space to velocity space in the rest frame of Mrk~817 at $z$ = 0.031455, using $z_{\textrm{outflow}}$. Figure~\ref{figVel2} shows this concept for visit 2n. To calculate the ionic column densities, we have chosen an integration range of $-3900$ to $-3600$ km~s$^{-1}$  (shown with vertical orange lines). 
This region was selected based on the centroid velocity and the width of the absorption trough of \civ. As all the absorption lines originate from the same outflow system, we employ the same integration range for all of them.  


To calculate the ionic column density of the \civ\  doublet, we employ the PC method and consider the results as a measurement. Since \siiv\  is shallow and weak, we use the AOD method and consider the results an upper limit. While \hi-\lya\ is actually a doublet due to its upper state having fine structure, it is treated as a singlet. The reason is that the energy levels of the fine structure are extremely close, so they cannot be separated spectroscopically.  For this reason, the column density of \lya\  is taken to be a lower limit, which is also consistent with saturation at full coverage of only the continuum (see Figure~\ref{figlya}), based on the AOD method. It is also blended with the ISM \nv\ 1238\AA\ absorption line (as shown in the fourth panel of Figure~\ref{figflux}). However, higher-order lines of the Lyman series are also covered in all \storm2 spectra, so we use Ly$\gamma$ as an upper limit for \hi\ column density. We do not use Ly$\beta$ since, as Figure~\ref{figflux} shows, it is contaminated by the \ovi\ broad absorption trough. The ionic column densities of  \nv, \ovi, and \svi\  were determined using the AOD method and are considered to be lower limits due to being saturated. \ciii's ionic column density is also measured using the AOD method and is considered a lower limit. While we only show the velocity plot for visit 2n, the same behavior was observed in visits 3n, 4d, and 75; hence, the same consideration will be applied to all visits. Table~\ref{tab2} provides the values of measured column density for each ion. The adopted uncertainties include the corresponding PC (for \civ) or AOD (for the rest of the lines) uncertainties and a systematic error of 10$\%$, added in quadrature \citep[e.g.,][]{xu18,mill18,mill20c,deh23}{}. The measurements reported for visit~3n slightly differ from the ionic column densities previously published in Paper~I. These minor differences are due to different measurement techniques, and the results are consistent within the uncertainties.

\begin{figure}
\centering
\begin{minipage}{3 in}
\includegraphics [width=3 in]{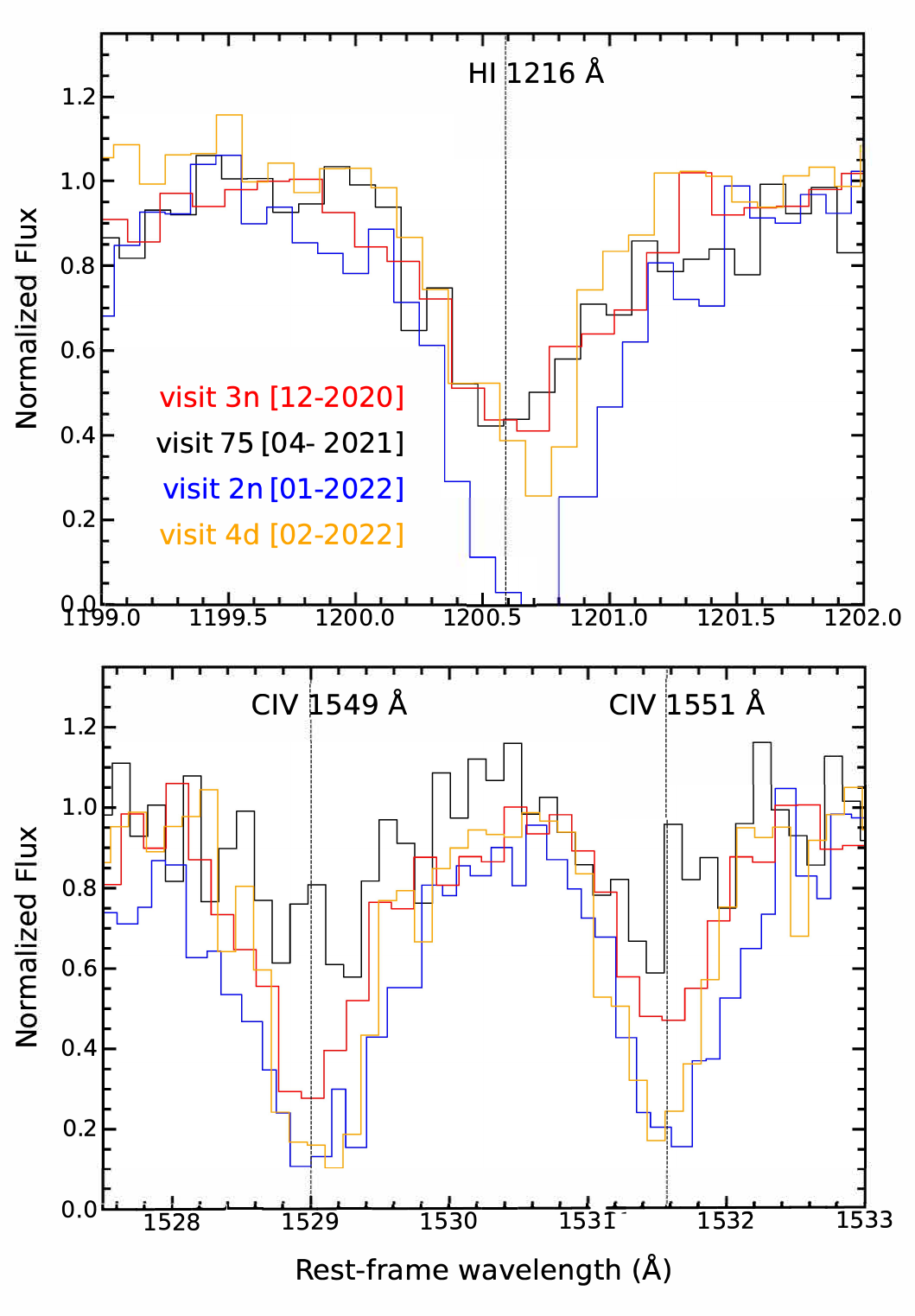}
\end{minipage}
 \caption{Top panel: The normalized flux in \lya \ absorption wavelength region from four \storm2 spectra. Bottom panel: The normalized flux in the \civ\  absorption wavelength region. The vertical dashed lines indicate the centroid wavelength of the absorption trough based on $z_{\rm outflow}$. Both panels show the normalized data after subtracting the emission model (see Figure~\ref{figlya}).}\label{figVary}
\end{figure}

While we investigate the properties of the outflow system and its possible variations later in Sections~\ref{sec:anal} and \ref{sec:dis}, it is worth mentioning that the depth of the absorption (the optical depth) varies from one visit to another, as shown for \civ\  in Figure 10 of Paper~I. 
Figure \ref{figVary} compares two examples of such variations (\lya \ and \civ\  absorption lines) by comparing the line profiles in detail. 

\begin{table}[h]
\caption{The ionic column densities, \storm2\label{tab2}}
 \centering
\setlength{\tabcolsep}{6pt} 
\renewcommand{\arraystretch}{1.3}
\begin{tabular}{c c c c c} 
 \hline
 Ion & Visit 3n& Visit 75 & Visit 2n & Visit 4d  \\ [0.6ex] 
 \hline\hline
\civ &  440$^{+60}_{-55}$& 340$^{+78}_{-69}$ &774$^{+220}_{-108}$ & 709$^{+140}_{-90}$   \\
 
 \siiv & <~21$^{+3}$ & <~12$^{+3}$&<~14$^{+3}$ &  <~7$^{+2}$ \\

 \hi-\lya &>~93$_{-10}$ & >~90$_{-11}$ & >~270$_{-75}$ &  >~100$_{-18}$ \\

 \hi-\lyg &<~900$^{+400}$ & <~1140$^{+600}$ & <~1200$^{+500}$ &  <~1500$^{+600}$ \\ 

 \nv & >~726$_{-85}$& >~490$_{-45}$ &>~900$_{-101}$ &  >~850$_{-104}$    \\
 
 \svi  & >~113$_{-45}$& >~112$_{-47}$ &>~116$_{-84}$ &  >~134$_{-53}$   \\

 \ovi & >~1808$_{-340}$& >~1500$_{-480}$&>~1080$_{-428}$  & >~1900$_{-455}$   \\
\ciii & >~45$_{-7}$& >~50$_{-20}$&>~54$_{-27}$  & >~59$_{-25}$ \\ 
\hline
\end{tabular}
\tablecomments{The column densities are in units of 10$^{12}$ cm$^{-2}$. We determine the lower limits using AOD measurements based on a Gaussian fit to the spectrum, accounting for blending and saturation. For upper limits, where absorptions are too shallow, we treat the AOD results as upper limits, adding a positive uncertainty. This approach ensures that we consider both finite lower [upper] and infinite upper [lower] uncertainties, as well as systematic errors.}
\end{table}

\subsubsection{Before \storm 2: visits 09-1 and 09-2}
In 2009, Mrk~817 was observed as a part of two non-\storm2 observing projects \citep{Pro09-1,pro09-2,wint11}. In both cases, the observations were obtained using COS with G130M and G160M gratings. These datasets are named visits 09-1 \citep{Pro09-1} and 09-2 \citep{pro09-2} in Table~\ref{tab1}, and both showed narrow \lya\ absorption lines in their spectrum. In both cases, since the \lya \ absorption is not saturated and is much shallower than the troughs observed in the \storm2 visits, we consider it as a measurement. For both of these visits, we could barely identify \civ\  and \nv\  absorption doublets. Given that these doublets are very shallow and are comparable to the level of the noise, we treat them as upper limits. These measurements are presented in Table~\ref{tab3} and will be subsequently employed in the photoionization modeling.

\begin{table}[h]
\caption{The ionic column densities, non-\storm\label{tab3}}
\centering
\renewcommand{\arraystretch}{1.3}
\begin{tabular}{c c c c} 
 \hline
 Ion & Visit 09-1 & Visit 09-2 & Visit A5 \\ [0.6ex] 
 \hline\hline
 \hi-\lya & 38$_{-4}^{+5}$ & ~42$_{-9}^{+9}$ & ~40$_{-11}^{+10}$ \\
\siiv & <5$^{+3}$  & <7$^{+3}$ & <~13$^{+7}$ \\
 \civ & <25$^{+5}$  & <27$^{+4}$ & <~35$^{+7}$ \\
 \ovi & - & - & >~460$_{-190}$ \\
 \svi & - & - & <~63$^{+23}$ \\
 \nv & <38$^{+6}$ & <45$^{+6}$ & 73$_{-5}^{+8}$ \\
 \hline
\end{tabular}
\tablecomments{The column densities are in units of 10$^{12}$ cm$^{-2}$.}
\end{table}
\subsubsection{After \storm2: Visit A5}
The most recent spectrum of Mrk~817 that we discuss here was obtained from HST observations of the source in July 2023 \citep{proptoo}. We identified the same absorption system in this spectrum through the \lya, \ovi\, and \nv\ narrow absorption lines. Since the weak \lya\  line is similar to the 2009 visits, we again take the estimates from the \lya \ as a measurement, while the column density of the \ovi \ doublet is measured via the AOD method and taken as a lower limit due to being saturated. We also identified the \nv\  doublet, and since it did not seem to be saturated, we took it as a measurement, too. We have also identified shallow troughs of \civ, \siiv, \ and \svi \ that can serve as upper limits. Table~\ref{tab3} reports the adopted value for each of the mentioned ionic column densities. 

\begin{figure}
\centering
\begin{minipage}{3.3 in}
\includegraphics [width=3.3 in]{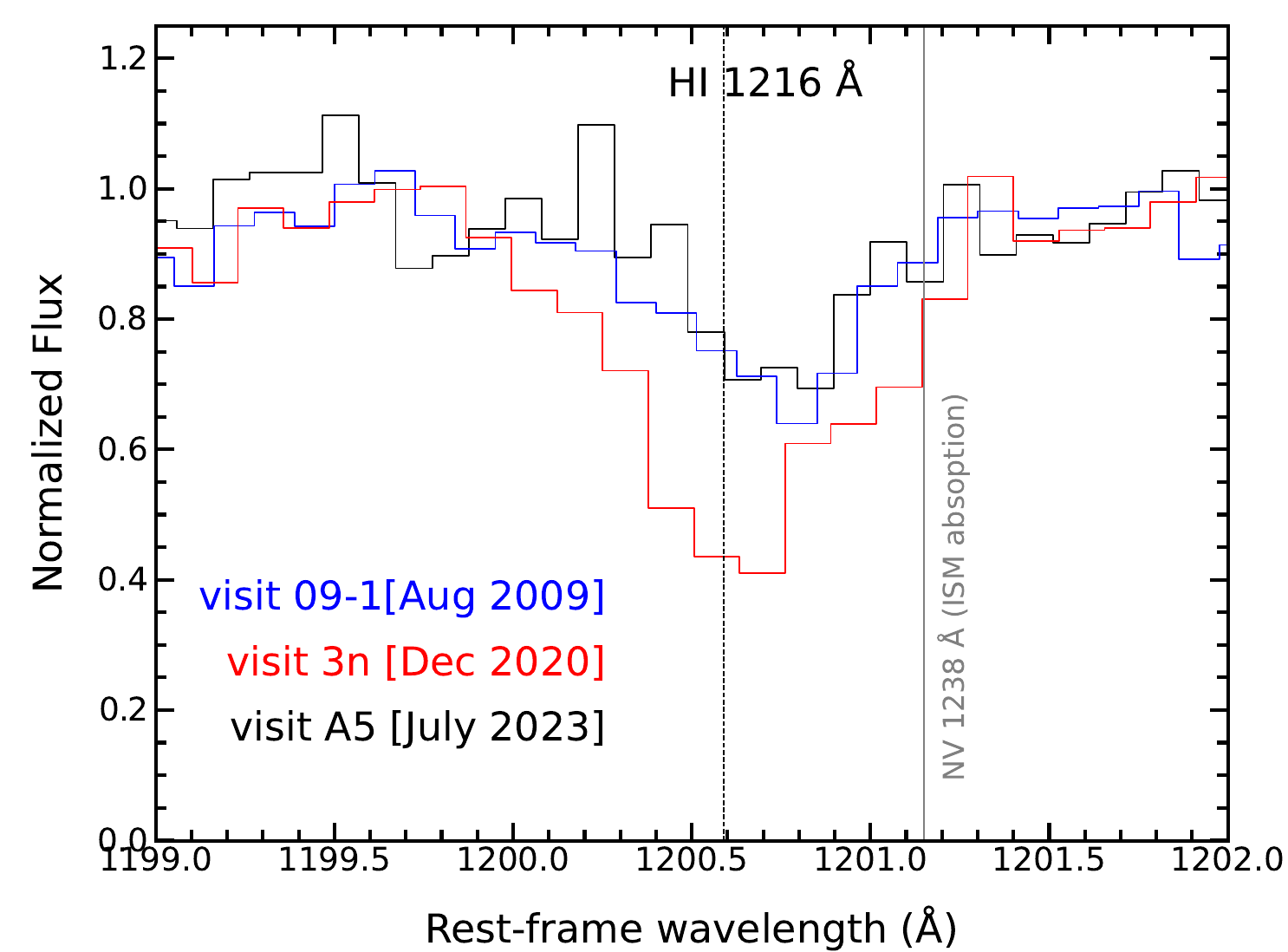}
\end{minipage}
 \caption{The \lya \ narrow absorption lines within visit 3n (\storm2) and visits 09-1 and A5 (both non-\storm2). The vertical dashed line indicates the centroid wavelength of the absorption trough based on $z_{\rm outflow}$. }\label{figVary-nonstorm}
\end{figure}

Figure~\ref{figVary-nonstorm} compares the \lya \ absorption troughs in one of the \storm2 visits (visit 3n) with one of the 2009 visits (visit 09-1) and the 2023 visit (visit A5). As illustrated in this Figure, the \lya \ observations from 2009 and 2023 exhibit similar depth, and both are shallower than the \lya \ absorption observed in December 2020 (visit 3n). This pattern 
supports the idea that AGN was in a higher ionization state in 2009 and 2023 compared to 2020.


\section{Photoionization Solutions}
\label{sec-photo}

The primary aim of this study is to estimate the properties of the absorption outflow system, including its total hydrogen column density ($N_{\rm H}$) and its 
ionization parameter ($U_{\rm H}$), and measuring the ionic column densities was the first step toward that goal \citep[e.g.,][]
{byun22a, byun22b, byun22c,deh23}.
The next step is to produce a grid of $N_{\rm H}$ and $U_{\rm H}$ using Cloudy simulations \citep{cloud23}. These simulations predict the ionic column density of each ion for each combination of $N_{\rm H}$ and $U_{\rm H}$. Cross-matching these predictions with the values deduced in Section~\ref{sec:anal} leads us to the outflow's $N_{\rm H}$ and $U_{\rm H}$.

We start this process by producing the appropriate SED for each visit. Because of the presence of the obscurer, for each visit we have two SEDs, an `unobscured SED' (i.e. the intrinsic continuum) that irradiates the obscurer, as well as the `obscured SED' that illuminates the further-away surrounding gas. As discussed (and shown) in Paper~I, the NALs outflow system is ionized by the obscured SED, implying that the obscurer is located between the outflow system and the source.

For each visit, we obtain the `unobscured' and the `obscured SED'  by
combining the results of Paper~I and 
\cite{part23} as follows: In Paper~I, the global model for the broadband continuum of Mrk~817 was established for one epoch (Visit 3n) using extensive HST and XMM-Newton observations. We
adopt the `warm Comptonization' version of this model in light of the

multi-wavelength variability characteristics of Mrk~817.
\begin{figure*}
\centering
\includegraphics [width=\textwidth]{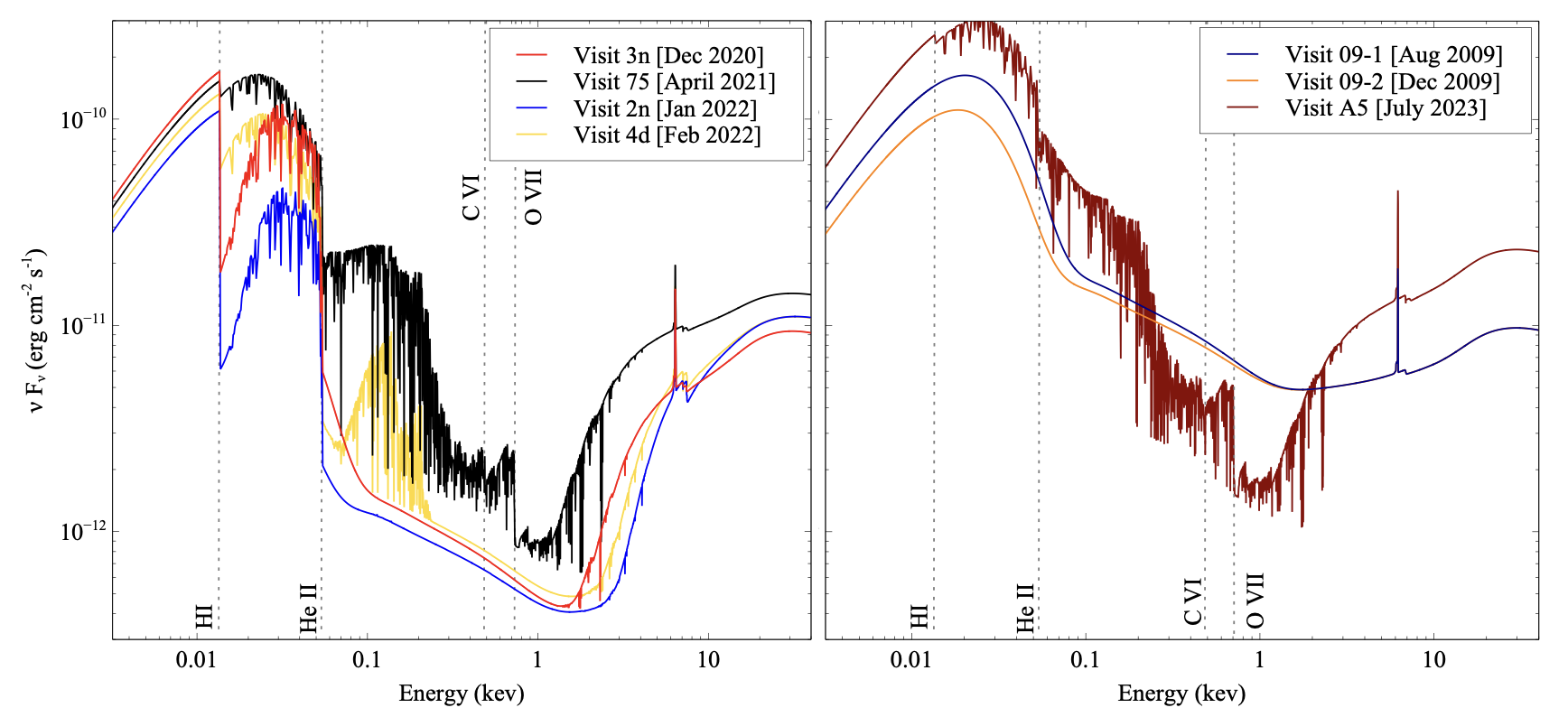}
 \caption{The SEDs used in the photoionization models (see Section~\ref{sec-photo} for  details). Left panel: The SED illuminating the narrow absorption outflow system during the \storm2 visits. Right panel: The SED illuminating the narrow absorption outflow system during non-\storm2 visits.  In both panels, the opacity sources of \hi\  (at 13.6 ev), \heii\  (at 54.4 ev), and the K-shell edges of carbon (\cvi\  at 489.99 ev), and oxygen (\ovii\  at 739.32 ev) are shown as dashed vertical lines.}\label{figsed}
\end{figure*}
In \cite{part23},
the parameters of the X-ray continuum and the obscurer (column density,
covering fraction, and ionization parameter) are derived for other
epochs using NICER monitoring observations. Therefore, by tracing
changes in the continuum and the obscurer parameters compared to the visit~3n, the unobscured and obscured SEDs for the other visits were
calculated. For each visit, the far-UV part of the broadband continuum
from Paper~I is matched to the observed HST spectrum by fitting the
temperature of the disk blackbody component of the SED. In the case of
the 2009 HST visits, which have no joint X-ray observations, the
X-ray part of the unobscured  SED from visit 3n is adopted. Figure~\ref{figsed} illustrates the SEDs used for each single visit. 

\begin{figure*}
\centering
\includegraphics [width=\textwidth]{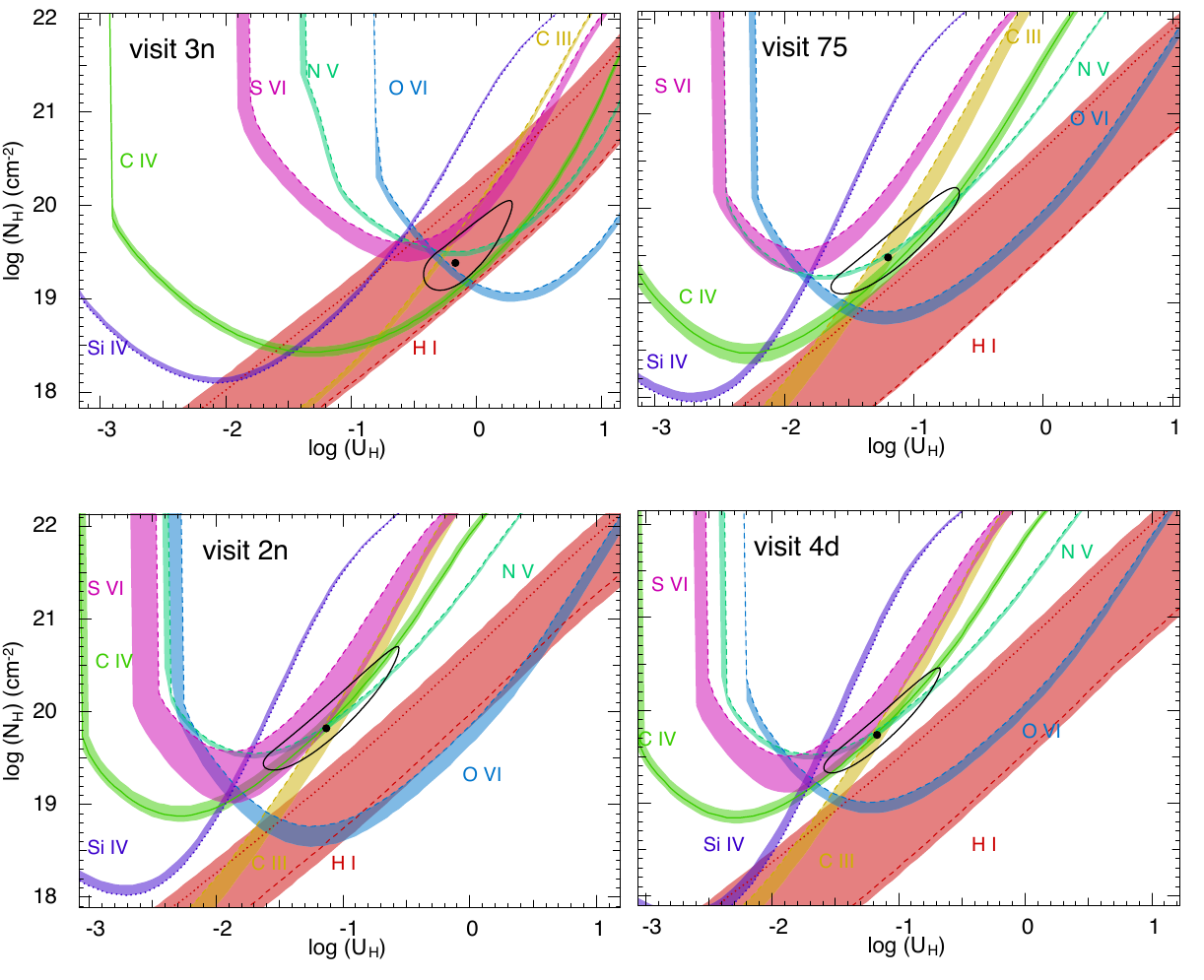}
 \caption{All four panels present single-phase photoionization solutions for the absorption outflow system per \storm2 visits. Each colored contour represents the ionic column densities consistent with the observations (refer to Table~\ref{tab2}), assuming the appropriate spectral energy distribution (SED) and solar metallicity. In all four panels, the \civ\  contour (green) consists of a solid line that indicates an actual measurement, while the upper and lower uncertainties form the contour's width. The dashed lines inside contours indicate that the estimated column density is indeed a lower limit. Dotted lines indicate the upper limits. Shaded bands depict the uncertainties added for each contour. The black circle denotes the best $\chi^{2}$-minimization solutions, and the 1~$\sigma$ confidence region is represented by a black contour.}\label{figNVUapp-storm}
\end{figure*}

\begin{figure}[h]
\centering
\begin{minipage}{3 in}
\includegraphics [width=3 in]{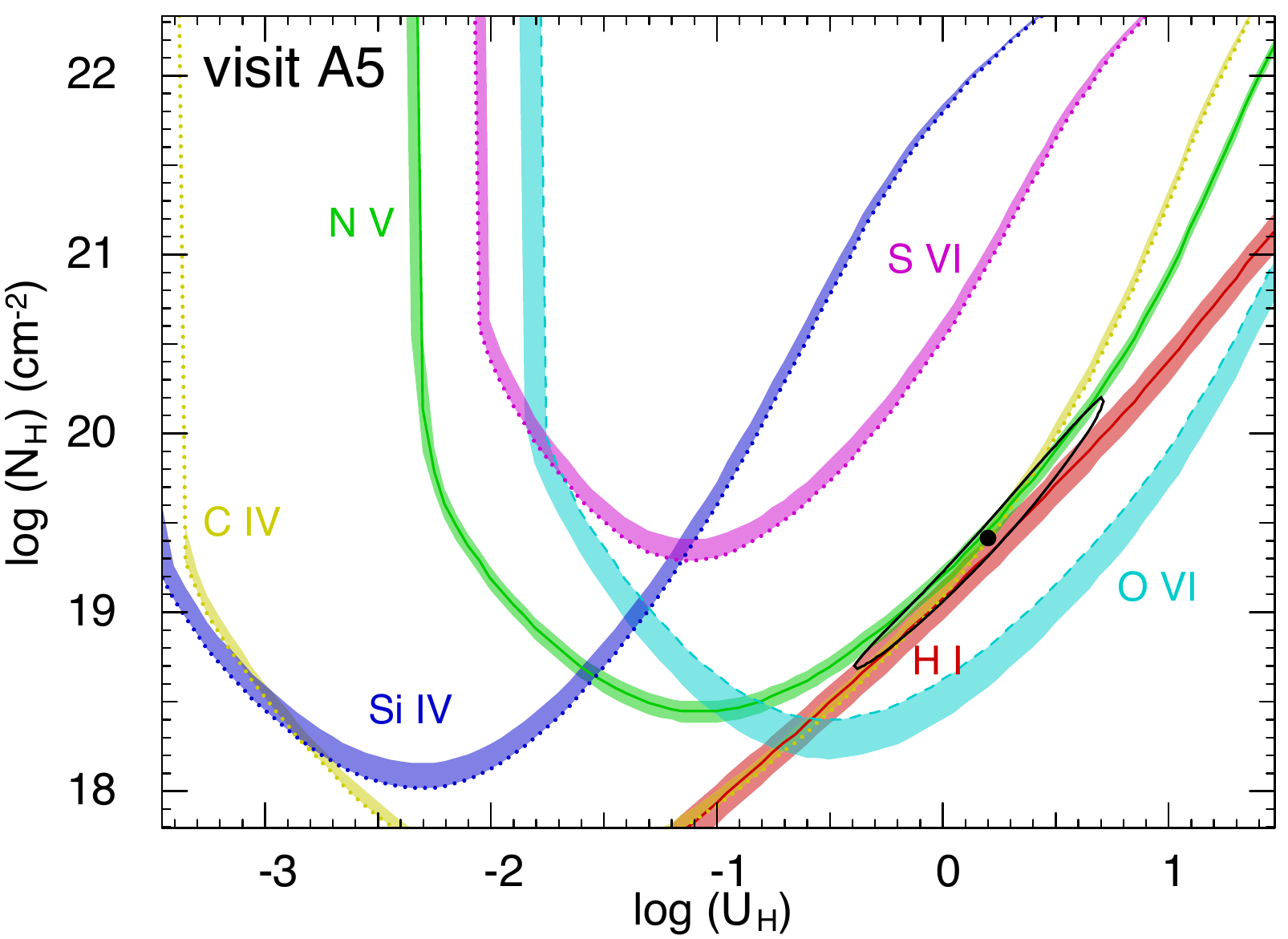}
\end{minipage}
 \caption{Same as Figure~\ref{figNVUapp-storm}, for non-\storm2 visit A5. Here we adopted the SED labelled as visit A5 in the left panel of Figure~\ref{figsed}.}\label{nvu-A5}
\end{figure}

With the required SEDs in hand, we produce Cloudy photoionization model grids to predict the characteristics of the outflow system. We use a fixed gas hydrogen density of $n_{\rm H}=10^{4}$~cm$^{-3}$ after verifying that the results are not dependent on the 
density for $n_{\rm H}<$ 10$^{11}$ cm$^{-3}$. Figure~\ref{figNVUapp-storm} displays the results of such calculations for \storm2 visits. Additionally, Figure~\ref{nvu-A5} presents the same results for the case of visit A5.
 
\begin{figure}[h]
\centering
\begin{minipage}{3 in}
\includegraphics [width=3 in]{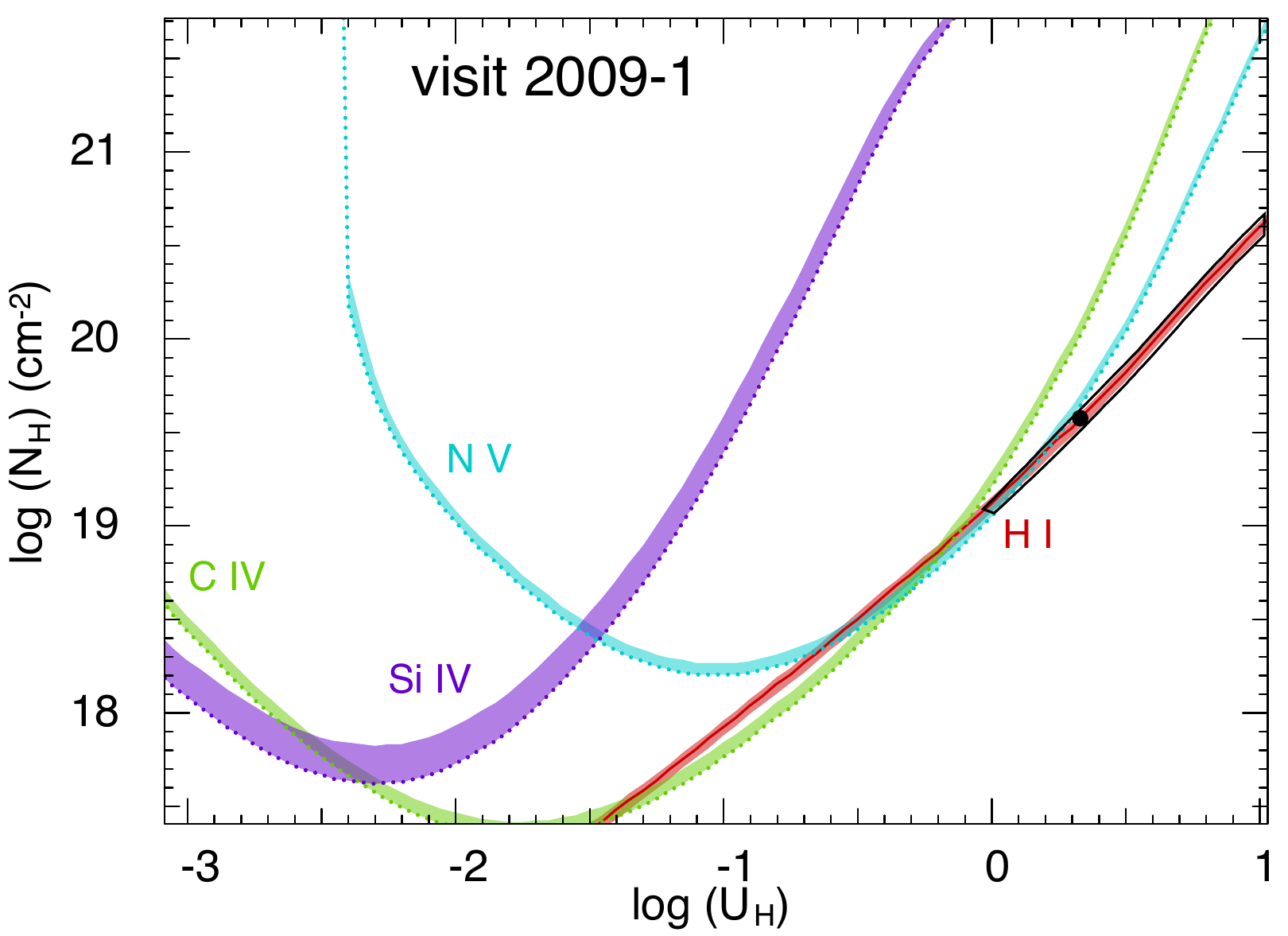}
\end{minipage}
 \caption{Same as Figure~\ref{figNVUapp-storm}, for non-\storm2 visit 2009-01. Here, we adopted the SED labeled as visit 09-1 in the left panel of Figure~\ref{figsed}.}\label{nvu-09-01}
\end{figure}

In each panel of Fig.~\ref{figNVUapp-storm}, ~\ref{nvu-A5}, and ~\ref{nvu-09-01} the colored contours match the values reported in 
Tables~\ref{tab2} and ~\ref{tab3}. We exclude visit 09-2 because the only measurement available for this visit is \lya\ while the upper limits implied by \civ, \nv, and \siiv\  are trivially satisfied. Therefore the photoionization solution can be anywhere along the \hi\  constraint. For each visit, and by employing $\chi^{2}$ minimization methods \citep{arav13}, we narrowed down the column density-ionization parameter space to a pair of 
$N_{\rm H}$ and $U_{\rm H}$ for the absorption outflow system. An absorption outflow system characterized by these $N_{\rm H}$ and $U_{\rm H}$ values yields the ionic column densities and their associated uncertainties given in Tables 2 and 3. The results of these simulations are detailed in Table \ref{tab4}. Note that solar abundances were assumed to produce these results. Due to the limited number of absorption lines identified in the two 2009 visits, which include only one measurement (\lya) and three upper limits (\civ, \siiv \ and \nv), there is considerable uncertainty associated with their $N_{\rm H}$ and $U_{\rm H}$.
\begin{table}[h]
\caption{Photoionization solution\label{tab4}}
 \centering
\setlength{\tabcolsep}{6pt} 
\renewcommand{\arraystretch}{1.3}
\begin{tabular}{c c c c} 
 \hline
 Visit  ID & $\log N_{\rm H}$ cm$^{-2}$ & $\log U_{\rm H}$ &$\rm Adj. ~\log U_{\rm H}$\\ [0.6ex] 
 \hline\hline
 09-1 & 19.55$^{+3.00}_{-0.50}$& 0.35$^{+1.50}_{-0.91}$&$-0.55^{+1.50}_{-0.91}$ \\
 \hline
 3n & 19.41$^{+0.70}_{-0.35}$& $-0.20^{+0.51}_{-0.60}$&$-0.70^{+0.51}_{-0.60}$ \\
 75 & 19.48$^{+0.74}_{-0.51}$& $-1.20^{+1.00}_{-0.42}$&$-1.90^{+1.00}_{-0.42}$\\
 2n & 19.82$^{+0.81}_{-0.45}$& $-1.13^{+0.90}_{-0.50}$&$-1.13^{+0.90}_{-0.50}$\\
 4d & 19.70$^{+0.71}_{-0.40}$& $-1.17^{+0.81}_{-0.41}$&$-1.67^{+0.81}_{-0.41}$\\
 \hline
A5 & 19.42$^{+0.70}_{-0.60}$& 0.20$^{+0.42}_{-0.81}$&$-0.88^{+0.42}_{-0.81}$\\
 \hline
\end{tabular}
\tablecomments{Adjusted $\log U_{\rm H}$ is a version of the ionization parameter that is scaled based on the variations of the SED. More details are available in Section~\ref{sec:dis}.}
\end{table}

\section{Discussion}
\label{sec:dis} 

\begin{figure}[h]
\begin{minipage}{3 in}
\includegraphics [width=3.5 in]{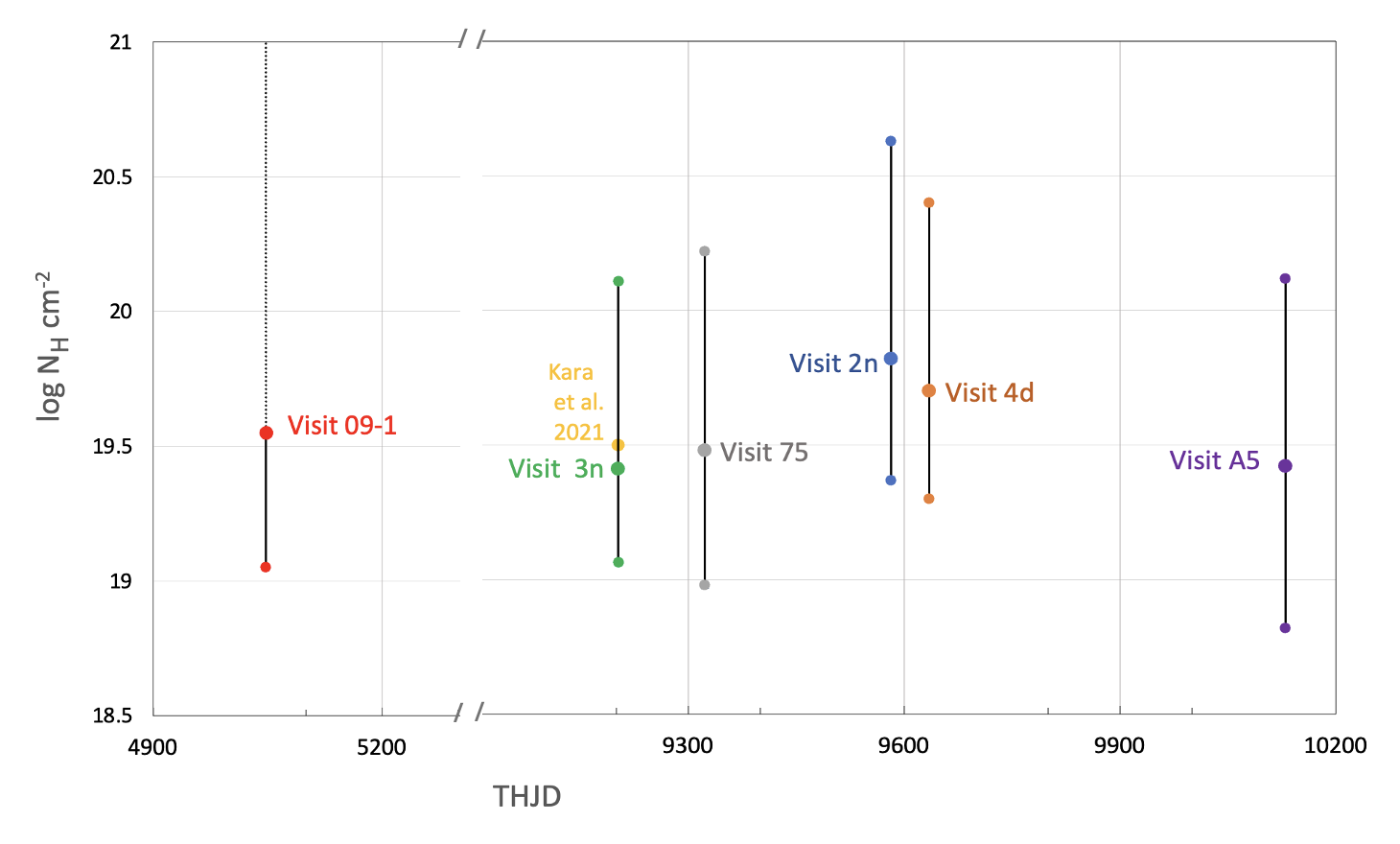}
\end{minipage}
 \caption{The hydrogen column density determined for each HST visit discussed here. We have also presented the column density measured by Paper~I for visit 3n. For each visit, the actual measurement is shown by a larger circle, while smaller circles indicate the upper and lower uncertainties. For visits 09-1 $\&$ 2, the dotted portion of the plot points to the large uncertainty in the value of the error.}\label{Ncompare}
\end{figure}

\begin{figure}
\begin{minipage}{3 in}
\includegraphics [width=3.5 in]{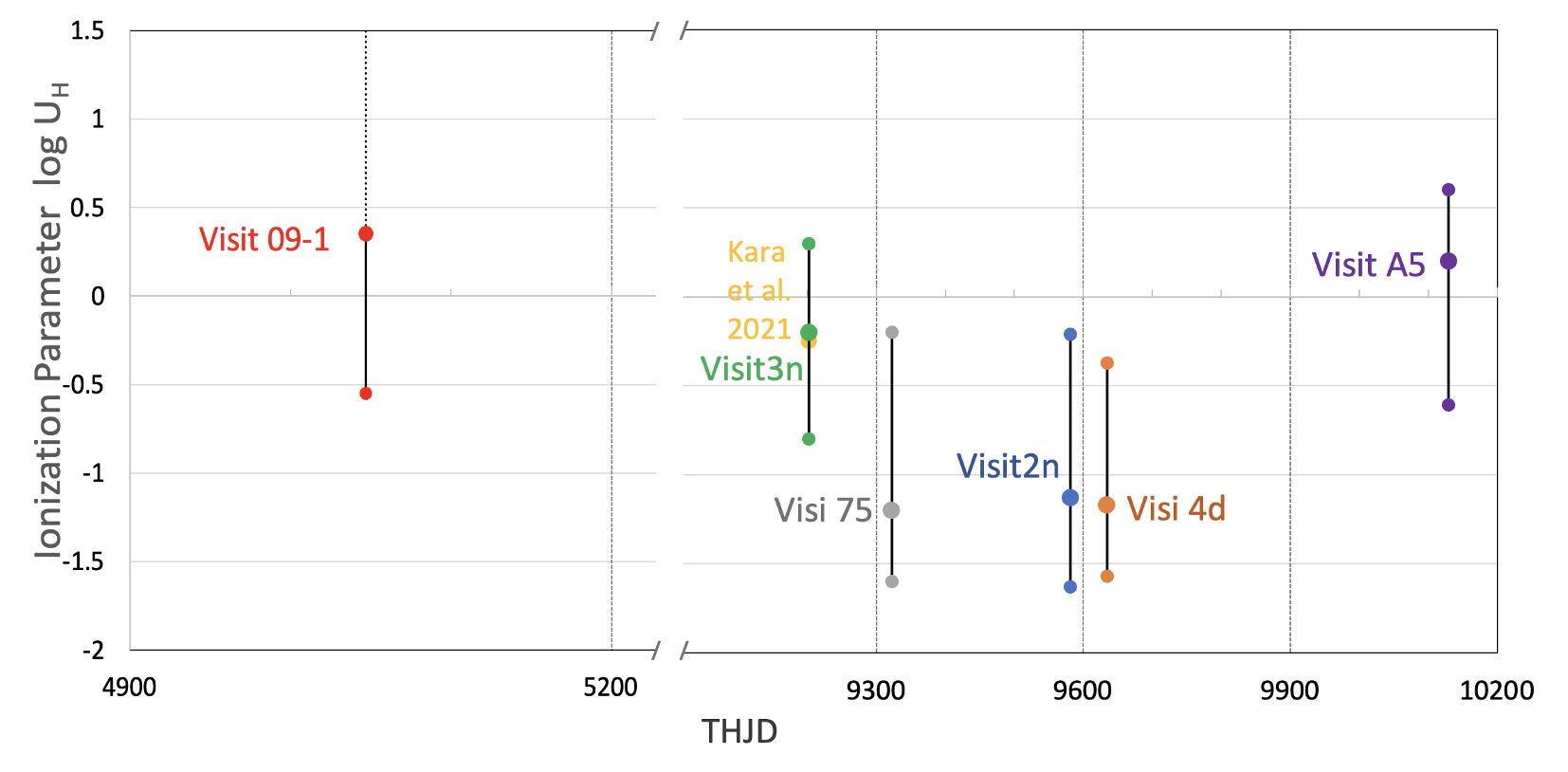}
\end{minipage}
 \caption{The ionization parameter $U_{\rm H}$ determined for each HST visit discussed here. We also present the ionization parameter measured by \cite{kara21} for visit 3n. For each visit, the actual measurement is shown by a larger circle, while smaller circles indicate the upper and lower uncertainties. For visit 09-1, the dotted portion of the plot points to the large uncertainty in the value of the error.}\label{Ucompare}
\end{figure}

\begin{figure}[h]
\begin{minipage}{3 in}
\includegraphics [width=3.5 in]{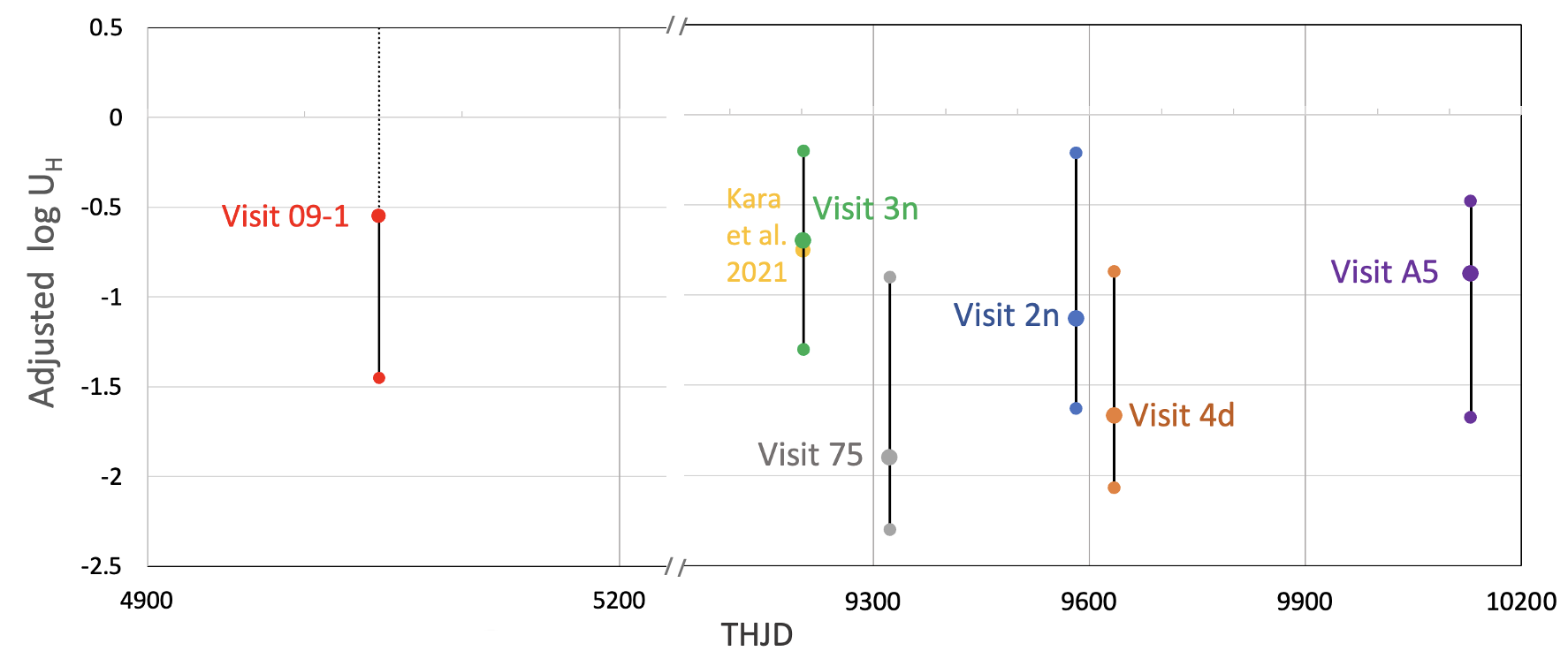}
\end{minipage}
 \caption{Similar to Figure~\ref{Ucompare}. Here the ionization parameters are adjusted with respect to the variations of the $Q({\rm H})$ (see text).}\label{adUcompare}
\end{figure}

Figures~\ref{Ncompare} to \ref{adUcompare} show how the measured column densities and two versions of the ionization parameter vary during 14 years.
Figure \ref{Ncompare} compares the total hydrogen column density of the absorption outflow system in each visit. Additionally, we include the measurements reported by Paper~I, consistent with our results.  The value of $N_{\rm H}$ is consistent between all epochs, supporting the assertion that this is the same stable outflow.

Figure~\ref{Ucompare} shows the ionization state of the absorption outflow system varies over time. We include the value of the ionization parameter measured by Paper~I \citep{kara21} (log $\xi$ = 1). Incorporating the best-fit values for the obscured SED in visit 3n, log$\xi$ = log U${_H}$+1.25. This implies that Paper~I's ionization parameter converts to log U${_H}=-0.25$, which is shown by the yellow circle in this figure. 

While the ionization parameter of the absorption outflow system is accurately measured for each visit, its value is affected by the AGN's flux variability as well as the variations of the obscurer from one visit to another. To understand this effect and to focus on the outflow's variations, it is crucial to consider only the integrated ionizing flux rather than only the observed UV flux. From the SED, We measure the $Q(\rm H)$, the number of hydrogen-ionizing photons
emitted by the central object per second \citep{oster06}. It is related to the ionization parameter by
\begin{equation}
U_{\rm H}\equiv\frac{Q(\rm H)}{4\pi R^{2}c~n_{\rm H}} 
\label{eq1}
\end{equation}
\noindent So for a constant hydrogen density and constant distance from the source: 
\begin{equation} 
\Rightarrow U_{\rm H} \propto Q({\rm H})\label{eq12}
\end{equation}
According to the equation above, we can adjust the ionization parameter using $Q({\rm H})$ for each visit. Figure~\ref{adUcompare} shows the results. To produce this figure, we first calculated the effects of SED variations on $U_{\rm H}$ (by calculating $Q(\rm H)$) and then scaled all ionization parameters with respect to one visit (visit 2n). This illustrates the variations of the outflow's ionization parameter, which are independent of the AGN's flux variability or the presence of the obscurer.

As depicted in Figure~\ref{adUcompare}, the measurements are consistent within the measured uncertainties. Drawing from these results and the outcomes of the column density calculations, we estimate that the outflow system possesses an adjusted ionization parameter log $U_{\rm H} =-1.0$ $^{+0.1}_{-0.3}$ and a total hydrogen column density of log $N_{\rm H}$ =19.5 $^{+0.61}_{-0.13}$ cm$^{-2}$. To estimate the uncertainties for log $N_{\rm H}$, we took into account the lower limit from visit2n and the upper limit from visit 3n. This guarantees that log $N_{\rm H}$ always works for all visits, even in its maximum or minimum limits. The same argument works for log $U_{\rm H}$ for which the upper and lower limits are based on visits 3n and 75, respectively.

As demonstrated by both Tables~\ref{tab2} and \ref{tab3}, the ionic column density derived for the \lya\  absorption line exhibited variability over a 14-year period. This variability suggests that the narrow absorption outflow system is associated with the AGN rather than being an intervening system. This serves as motivation to determine the location of the NAL outflow system. Due to the absence of density-sensitive, excited-state narrow absorption lines in the examined HST visits, we were unable to pursue the methods presented in \cite{byun22b, byun22c} and \cite{deh23} for determining the location.  However, following the discussion in \cite{arav12}, we can obtain limits on the electron density based on the limits on the observed recombination time, which then results in an upper limit on the distance for the outflowing gas. We use their discussion to estimate the electron density of the outflow based on the \lya\  variability during a certain period of time:

\begin{equation}
 \begin{multlined}
t^* = \left[ -f \alpha_i n_e \left( \frac{n_{i+1}}{n_i} - \frac{\alpha_{i-1}}{\alpha_i} \right) \right]^{-1}, 
\label{eq-rMax}
 \end{multlined}
\end{equation}
\label{eq-rMax}
\noindent in which  $t^{*}$ is the timescale for changes in the ionic fraction, $f$ is a scaling factor or ionic fraction, $\alpha_{i}$ is the recombination rate coefficient for the ionization state $i$. $n_{e}$ is the electron density, while $\frac{n_{i+1}}{n_i}$ is the ratio of the number densities of the next ionization state ($i+1$) to the current state ($i$). And finally, $\frac{\alpha_{i-1}}{\alpha_i} $ is the ratio of recombination coefficients between the ionization state ($i-1$) and the current state ($i$).

We set limits on the recombination time, $t^{*}$, by considering
the shortest time span between two visits that exhibit unambiguous changes in \lya\  absorption. Since \storm2 observations
show significant \lya\  absorption variations between adjacent epochs separated by two days, we determine an upper limit on the recombination time of two days. However, it's plausible that the actual recombination time is smaller, so we treat the derived density as a lower limit, resulting in an upper limit for the location, denoted as $r_{\rm max}$.

To proceed with our calculations, Cloudy 23.0 \citep{cloud23} was employed to determine the ratio of $\frac{n_{i+1}}{n_i}$ for a narrow absorption outflow system that is ionized by the SED corresponding to visit 2n. Following the discussion in \citep{krolik95} and assuming equilbrium, we can take $f$=$-1$. Consequently, our analysis yields a calculated upper limit for the electron density to be $log~n_{\rm e} > $3.46$\,{\rm cm}^{-3}$ or $log~n_{\rm H} > $3.5$\,{\rm cm}^{-3}$. We then solve Equation~\ref{eq1} to compute the corresponding location, resulting in a value of $r_{\rm max} =38~ \rm pc$ for the distance between the NAL outflow system and the central source.


As discussed in Paper~I and also in Section~\ref{stormanal} here, resolving the peculiar depth ratios of the \nv, \civ, and \ovi\  doublet troughs is achievable if the narrow-line absorption only covers the continuum source.  Paper~I speculated that this suggests the NAL outflow system is located within the interior of the BLR. Alternatively, another possible scenario 
is that the narrow absorbing cloud is relatively small, sufficiently covering the continuum-emitting region ($\sim$100 $R_{\rm 
g}\approx$5.7$\times$10$^{14}$cm, for a black hole mass of 3.85$\times 10^{7}$ \(M_\odot\)  \citep{bent15} but not large enough to encompass a substantial portion of the BLR ($\sim$10 light days or 2.6 $\times$10$^{16}$cm \citep{kara21,hom23a}). Its low hydrogen density (>3000 cm$^{-3}$) and a column density of 3$\times$10$^{19}$ cm$^{-2}$ suggest a maximum thickness of 1$\times$10$^{16}$ cm for the cloud. Thus, regarding its small size, it could take on a quite elongated "string" shape with an aspect ratio of $\sim$15:1
. Its high velocity and possible elongated morphology might suggest it is material that has been entrained in a faster outflow, perhaps one of the ultra-fast outflows (UFOs) identified by \cite{zak23}. However, if the cloud instead has a density of 3$\times$10$^{5}$ cm$^{-3}$ (typical of the NLR is other Seyferts like NGC 5548, e.g., \cite{arav15}), which locates it much closer to the source (3.8 pc), it would be consistent with being spherical.

If the cloud is small in diameter, the fact that its properties in absorbing the continuum are stable suggests that its transverse motion must be small. To put this into perspective, considering a cloud with a diameter of 5.7$\times10^{14}$ cm, for it to traverse this distance in less than 14 years, its transverse velocity would need to be approximately 13 km~s$^{-1}$. In contrast, the Keplerian velocity around the black hole in Mrk~817 (with a mass of $3.85 \times 10^7$ \(M_\odot\)) ranges between 209 and 66 km~s$^{-1}$ for radii spanning from 3.8 to 38 pc. Although the limit of 13 km~s$^{-1}$ falls below these estimates, clouds with several times larger diameters could still effectively obscure the continuum without substantially covering the BLR.

The high-velocity narrow absorption lines in Mrk~817 have become noticeable specifically during the epochs in which the X-ray obscurer exists.
This phenomenon is observed in various cases, such as Component~1 in NGC 5548 \citep{arav15} and the stronger \lya\  line witnessed in the recent obscuring event in MR2251-178 \citep{mao22}. Notably, these lines were present previously but gained prominence only during the obscuration of the ionizing continuum. This suggests that the observed gas normally exists in a fairly high ionization state with only \lya\  and other high-ionization lines weakly visible at all. Their presence becomes more prominent when the ionizing continuum is significantly diminished due to obscuration, lowering the ionization state of the gas and making lower ionization species more prominent. These large-scale outflows could originate from a torus \citep[e.g.,][]{Dorod08, Dorod16}, an outer accretion disk  
\citep[e.g.,][]{wat21} or from inflows
\citep[e.g.,][]{Prog07, Kuro09, Mocib13}.

\section{Summary}
\label{sec:sum} 
In this paper, we examined the narrow absorption outflow system in seven distinct spectral epochs of Mrk~817, all observed by HST between 2009 and 2023. We identified several narrow absorption lines in each visit and subsequently measured the ionic column densities, which were later employed for photoionization modeling purposes. The detailed results are presented in Table~\ref{tab4}.
Our analysis of the absorption outflow system in Mrk~817, spanning seven separate HST spectra from 2009 to 2023, has provided insights into the system's stability, i.e., consistent $N_{\rm H}$ and $U_{\rm H}$. We summarize our results as follows: 

\begin{itemize}
    \item We have identified the same high-velocity narrow absorption line outflow system in all seven HST visits spanning over 14 years.
    
    \item The narrow absorption outflow system is ionized by the "obscured" SED, confirming that the obscurer is between the narrow absorption cloud and the central source.

     \item Based on our findings, we estimate that the outflow system has an `adjusted' ionization parameter log $U_{\rm H} =-1.0$ $^{+0.1}_{-0.3}$ and a total hydrogen column density of log $N_{\rm H}$ =19.5 $^{+0.61}_{-0.13}$ cm$^{-2}$.
    We determine that this outflow system is connected to the AGN and is situated at a distance of <38~pc from the central source. It also has a hydrogen gas number density whose value exceeds 3000 cm$^{-3}$.
    
    \item The observed consistency in $N_{\rm H}$ across all visits and variations in $U_{\rm H}$ that are in concert with a response to changes in the ionizing continuum suggest that the outflow system has been persistent through out the a 14-year period.

\end{itemize}

\facility{HST(COS, STIS)}
\software {We used the Python astronomy
package Astropy \citep{astro13,astro18} for our
cosmological calculations, as well as Scipy \citep{virt20},
Numpy \citep{harr20}, and Pandas \citep{reba21} for most of our numerical computations. For our plotting purposes, we used Matplotlib \citep{hunt07}.}
 \section*{Acknowledgements}
We express our appreciation to the anonymous reviewer whose feedback has contributed to the enhancement of this manuscript.
All of the data presented in this paper were obtained from the Mikulski Archive for Space Telescopes (MAST) at the Space Telescope Science Institute. The specific observations analyzed can be accessed via \dataset[10.17909/16n2-nt70]{https://doi.org/10.17909/16n2-nt70}.
Our project began with the successful Cycle 28 HST
proposal 16196 \citep{pet20}. M.D, N.A, D.B, G.W, and M.Sh acknowledge support
from NSF grant AST 2106249, as well as NASA STScI grants AR-
15786, AR-16600, AR-16601, and HST-AR-17556. E.M.C. gratefully
acknowledges support from NASA through grant 80NSSC22
K0089. Y.H. acknowledges support from the Hubble Space
Telescope program GO-16196, provided by NASA through a
grant from the Space Telescope Science Institute, which is
operated by the Association of Universities for Research in
Astronomy, Inc., under NASA contract NAS5-26555.
C.S.K. is supported by NSF
grant AST-2307385. M.C.B. gratefully acknowledges support
from the NSF through grant AST-2009230. H.L.
acknowledges a Daphne Jackson Fellowship sponsored by the Science and Technology Facilities Council (STFC), UK.
D.I., A.B.K, and L. \v C.P. acknowledge funding provided by the University of Belgrade—Faculty of Mathematics
(contract 451-03-66/2024-03/200104), Astronomical Observatory Belgrade (contract 451-03-66/2024-03/200002),
through grants by the Ministry of Education, Science, and Technological Development of the Republic of Serbia. D.I.
acknowledges the support of the Alexander von Humboldt Foundation. A.B.K. and L. \v C.P. thank the support by Chinese
Academy of Sciences President’s International Fellowship Initiative (PIFI) for visiting scientists. 
A.V.F. is grateful for
financial assistance from the Christopher R. Redlich Fund
and numerous individual donors. M.V. gratefully acknowledges
support from the Independent Research Fund Denmark
via grant No. DFF 8021-00130.  CSK is supported by NSF grants AST-2307385
and AST-1908570. M.R.S. is supported by the STScI Postdoctoral Fellowship. PK acknowledges support from NASA through the NASA Hubble Fellowship grant HST-HF2-51534.001-A awarded by the Space Telescope Science Institute, which is operated by the Association of Universities for Research in Astronomy, Incorporated, under NASA contract NAS5-26555. J.G. gratefully acknowledges support from NASA through the grant 80NSSC22K1492.

 \end{document}